\documentclass[11pt,twoside,letterpaper]{article}
\usepackage[utf8]{inputenc}
\usepackage[T1]{fontenc}
\usepackage{amsmath}
\usepackage{amsfonts}
\usepackage{amssymb}
\usepackage{xcolor}
\usepackage{mathtools}
\usepackage[title]{appendix}%
\usepackage[export]{adjustbox}%
\usepackage{algorithm}%
\usepackage{enumerate}
\usepackage{hyperref}
\usepackage{doi}
\usepackage{array}
\newcolumntype{s}{>{\footnotesize}l}
\usepackage{booktabs}
\hypersetup{pdfborder={0 0 0},
	colorlinks=true,
	linkcolor=black,
	citecolor=black,
	urlcolor=black}
\usepackage{fourier}
\usepackage{pifont}
\usepackage{orcidlink}
\usepackage[font=small]{caption}
\newcommand{\cmark}{\ding{51}}%
\newcommand{\xmark}{\ding{55}}%

\date{Compiled \today}

\usepackage{verbatim}
\usepackage{longtable}
\DeclareMathOperator{\diag}{diag}
\usepackage[left=3.2cm, right=3.2cm]{geometry}
\usepackage[calcwidth,  sf,  big, compact]{titlesec}
\titleformat{\section}
{\normalfont\scshape \Large}{\thesection}{1em}{}
\usepackage{fancyhdr}
\usepackage{setspace}

\usepackage[authoryear]{natbib}
\usepackage{cleveref}
\renewcommand{\d}{\mathrm{d} }
\newcommand{\I}[2][]{{\mathbf 1}_{#1}\left(#2\right)}
\titleformat{\section}[block]{\large \bf }
{  {\thesection.}}{4pt}{   }
\titleformat{\subsection}[block]{\itshape}
{  {\thesubsection.}}{4pt}{   }
\titleformat{\subsubsection}[block]{\itshape}
{  {\thesubsection.}}{4pt}{   }

\title{An utopic adventure in the modelling of conditional univariate and multivariate extremes}
\author{L\'eo R. Belzile\thanks{Department of Decision Sciences, HEC Montréal (\texttt{leo.belzile@hec.ca})}, Arnab Hazra\thanks{Department of Mathematics and Statistics, Indian Institute of Technology Kanpur}\ and Rishikesh Yadav\thanks{Department of Decision Sciences, HEC Montréal}
}
\date{}

\begin{document}
\maketitle

\begin{abstract}
 The EVA 2023 data competition consisted of four challenges, ranging from interval estimation for very high quantiles of univariate extremes conditional on covariates, point estimation of unconditional return levels under a custom loss function, to estimation of the probabilities of tail events for low and high-dimensional multivariate data. We tackle these tasks by revisiting the current and existing literature on conditional univariate and multivariate extremes. We propose new cross-validation methods for covariate-dependent models, validation metrics for exchangeable multivariate models, formulae for the joint probability of exceedance for multivariate generalized Pareto vectors and a composition sampling algorithm for generating multivariate tail events for the latter.  We highlight overarching themes ranging from model validation at extremely high quantile levels to building custom estimation strategies that leverage model assumptions.
\end{abstract}
\section{Introduction}

The data competition of the 2023 edition of the Extreme Value Analysis Conference (EVA 2023) assigned a series of four challenges to participants.  In designing these, the organizers sought to capture end-user considerations that applied statisticians are faced with, notably estimating quantiles and probabilities of extreme events in univariate and multivariate settings. The exercise uses simulated data sets representing the behaviour of certain environmental parameters in the hypothetical country `Utopia'. Further details regarding the competition can be found in \cite{rohrbeck2023editorial}.

The first two challenges focus on conditional univariate extreme value analysis, where the objective of the first task is interval estimation for very high conditional quantiles and that of the second is the estimation of 200-year unconditional return level integration over the density of the covariates, using a customized loss function that penalizes underestimation more compared to overestimation. Both challenges used the same data set representing the equivalent of daily measurements for 70 years.

Compared to the independent and identically distributed setting, there is no overarching theory for conditional univariate extremes modelling. There are three main approaches for handling dependence on covariates: the first one assumes the parameters of the models for univariate extremes to be certain functions of the explanatory variables. Early work on regression models for extremes \citep{smith1989extreme, Davison.Smith:1990} assumed simple regression settings for parameters of the extreme value distributions, with extensions based on local likelihood \citep{Davison.Ramesh:2000,hall2000nonparametric} or penalized likelihood \citep{pauli2001penalized}, etc. Instead of a simple regression setting, \cite{chavez2005generalized}, \cite{yee2007vector}, and \cite{Youngman:2019} considered a generalized additive model \citep[GAM,][]{Hastie.Tibshirani:1986} framework that assumes the parameters of the generalized Pareto distribution to be some unknown but smooth functions of the covariates. The second avenue is the use of nonstationarity thresholds \citep{Northrop:2011, Youngman:2019}, modelled using quantile regression or otherwise. A third option, which we did not explore, is the modelling of residuals from a model fitting to the bulk of the observations \citep{Eastoe:2009}, as trends are easier to detect with more observations. More recent proposals combine machine learning methods for statistical learning with extreme value theory through local likelihood with extremal random forests \citep{gnecco2022extremal} and gradient boosting \citep{Velthoen:2023}.

Most tasks involve extrapolation, and thus one must wonder whether the latter is trustworthy. To answer this question, we need goodness-of-fit assessment, model validation, and model comparison tools. Since rare events are scarce and our targets lie in estimating or predicting much beyond the range of the observations, we rely on threshold stability by validating the model at observed levels. Diagnostic tools such as quantile-quantile plots can be adapted to the case of non-identically distributed data \citep{Davison.Smith:1990}, but the model comparison is complicated by the fact that the models for the threshold exceedances do not feature the same data unless we fix the threshold for all competing models. Such issues are exacerbated when the threshold varies as a function of covariates. Few works focus on validation; likelihood-based inference allows for the use of information criteria and tests for nested data, while more generally scoring rules \citep{Gneiting.Raftery:2007} can be used. Threshold-weighted scoring rules \citep{Gneiting.Ranjan:2011} can be used to give more weight to extreme events and have been employed in extremes \citep[e.g.,][]{Huser:2021}, but this can lead to paradoxes \citep{Lerch:2017}. There is little work on cross-validation for univariate peaks over threshold models; among these, \cite{Northrop:2017} uses cross-validation for threshold models in a Bayesian framework, whereas \cite{Gandy:2022} proposes to validate at much lower levels, appealing to threshold stability for extrapolation. There is no direct extension of the available approaches to the covariate-dependent cases. As a novel contribution, we propose a cross-validation algorithm (\Cref{algo:intervalscore}) for interval estimation for the covariate-dependent models.

Custom loss functions and evaluations of credible intervals for predictive inference for extremes are discussed in \cite{Smith:1999}. We approximate the loss function pointwise by averaging the loss over the posterior samples to obtain the return level estimate and discuss the effect of the choice of the loss function on the final inference, for more details refer to \Cref{sec:partC2}.

Estimation of multivariate probability of exceedances is challenging in high dimensions. Early semiparametric approaches for extrapolation were built on regular variation \citep{deHaan.Resnick:1977} and were extended in \cite{ledford1997modelling}, by building structure variables \citep{Coles.Tawn:1994} and estimating the tail index of the latter. \cite{Wadsworth.Tawn:2013} generalized the method for angles in different directions than the origin. While parametric models for multivariate extremes have existed since the 1990s \citep[cf.][]{Coles.Tawn:1991} under the assumption of max-stability and asymptotically dependent limits, regression modelling for asymptotically independent extremes took off with the conditional extreme value model of \cite{Heffernan.Tawn:2004}, further generalized in \cite{Keef:2013,Keef:2013b}. In the presence of covariates, the parameters of the conditional extreme value model can be assumed to be dependent on them \citep{Jonathan.Ewans.Randell:2013}. Many of these approaches can be related through the notion of geometric extremes \citep{Nolde.Wadsworth:2022}, for which statistical inference is still in its infancy \citep{Wadsworth.Campbell:2022}. While the mentioned models are flexible and theoretically justified, many of them grapple with the curse of dimensionality since extrapolation frequently involves simulation from the empirical distribution. However, these models can be further simplified through additional data exploration and with the tasks in hand, as demonstrated in the third and fourth tasks of the data competition; see \Cref{sec:partC3,sec:partC4} for a more detailed explanation.

The third and fourth tasks focused on the estimation of the probabilities of joint exceedances when all or some of the variables are large, with again potential dependence on covariates in Task~3. Data have known marginal distributions, so the focus is solely on the dependence structure. More specifically, the third task focuses on the simultaneous exceedance of a fixed high level for all components of a trivariate random vector with known marginal distributions while the other part targets estimating the probability of the simultaneous exceedance of an even higher level for only two components when the remaining component is smaller than its median.

The fourth task aims to estimate the probability of jointly exceeding certain marginal quantile levels for a 50-dimensional random vector. This naturally requires dimension reduction methods for model fitting, as multivariate extreme value analysis is extremely challenging if not infeasible in such dimensions. An exploratory data analysis reveals the presence of clusters of exchangeable components, a structure we leverage to reduce the complexity of the problem. This allows us to simplify the models to a great extent. Because of the infinitesimally small values of the joint tail probabilities, maintaining numerical stability is also necessary and we also adapt the tail probability estimation using numerical methods to bypass the challenges raised by events that have probabilities so small that Monte Carlo methods are inaccurate.

The paper is organized as follows: we present each task assigned in turn, discussing briefly the data and the challenges. We describe the methodology required to solve such problems, highlight the somewhat pragmatic choices we made due to time constraints, and conclude with a postmortem of the results and an assessment of our performance. The closing section discusses broader implications for applied projects, lists our main contributions, and contains a reflection on lessons learned by partaking in the data challenge.

\section{\texorpdfstring{Task~1: Estimating confidence intervals for the extreme conditional quantiles}{Task C1}}

\subsection{Data and task description}
 We have access to 21000 training observations, available over 70 years with each year comprising 300 days. There are twelve months in a year with each month comprising 25 days. There are two seasons each of 150 days; the first and the last six months represent two different seasons.
The data contains a response variable, $Y$, and eight covariates: $V_1$, $V_2$, $V_3$, $V_4$, Wind direction, Wind speed, Season, and Atmosphere. Most explanatories are independent and identically distributed marginally, except for $V_3$ whose marginal distribution depends on Season, and for Atmosphere which is constant within months, but cyclical over 70 years. Some covariates have data known to be missing completely at random (MCAR): 11.69\% of the training observations have missing data for at least one covariate. Among the individual variables, the marginal percentages of missing data in $V_1$, $V_2$, $V_3$, $V_4$, Wind direction, and Wind speed are 2.05\%, 1.98\%, 1.90\%, 2.17\%, 2.04\%, and 2.15\%, respectively.

In Task~$1$, our objective is to provide 50\% confidence intervals for the 0.9999{th} conditional quantiles for 100 different levels of the covariates from a validation set. After the competition is over, the organizers calculate the percentage of cases where the submitted intervals cover the true 0.9999{th} conditional quantiles were known to the organizers up to some negligible Monte Carlo error. A team with a coverage percentage closer to 50\% was assigned a better ranking in the sub-competition; in the case of ties for the coverage, the team with smaller average interval lengths was favoured.

\subsection{Exploratory data analysis}

\begin{figure}[htbp!]
 \centering
 \includegraphics[width = 0.9\textwidth]{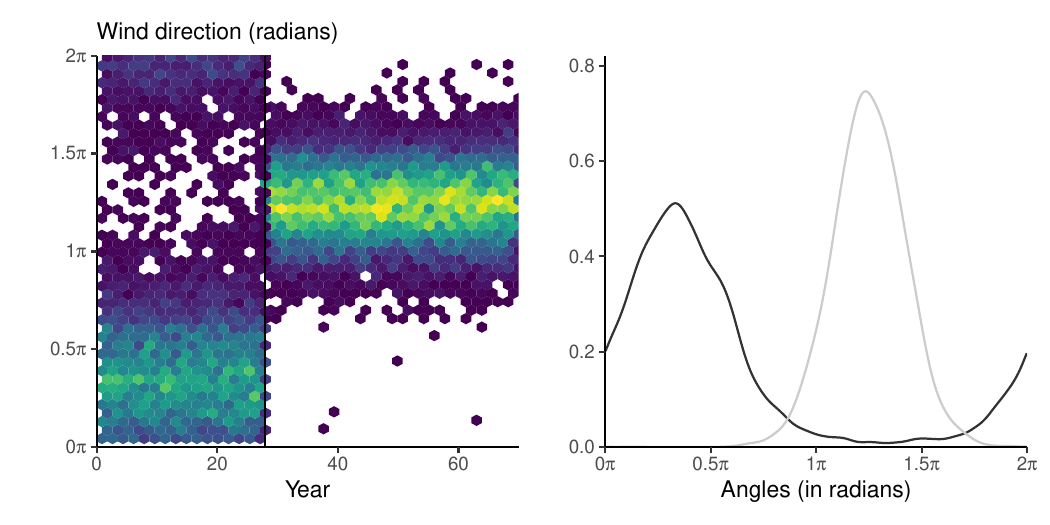}
 \caption{Kernel density estimator of Wind direction (left) as a function of the observation index, and the circular density estimates per cluster (right). We binned Wind direction per year; lighter shades for hexagonal bins indicate higher density.}
 \label{fig:angles}
\end{figure}

An artifact of the data creation is that there are two distinguishable clusters for Wind direction with a clear change in the distribution occurring circa observation number 8357, as shown in the left panel of \Cref{fig:angles}: the organizers wanted Wind direction to be independently drawn from a mixture of two components, but forgot to randomize the series. The realizations of Wind direction are split between two modes containing roughly 60\% and 40\% of the series, with dominant mean wind direction of 225$^{\circ}$ and 60$^{\circ}$, respectively. Abstracting from the circular nature of wind direction, we fit a changepoint algorithm under the assumption of normality to estimate the index at which this mean-variance change occurs (the structural break index could also be identified manually). We create a binary explanatory variable for wind regime, labeled `Changepoint' henceforth, and use a circular kernel density model to estimate the direction in Task~$1$, shown in the right panel of \Cref{fig:angles}: we use the resulting density estimator to predict the class of the 100 holdout covariate sets.

Spearman's rank correlation matrix (not shown) suggests that, outside of the pairs of covariates ($V_1, V_2$) and (Wind Speed, Wind Direction), variables are uncorrelated. The challenge description stated that the marginal distribution of $V_3$ depended on Season. We used energy tests of independence \citep{Rizzo.Szekely:2010} to check whether covariates are independent and found no evidence against this hypothesis for the following tuples: ($V_1, V_2$), ($V_3$, Season), $V_4$, (Wind Speed, Wind Direction, Atmosphere). One could exploit this knowledge to impute missing data points and avoid spurious regressions, or to average over the covariate distribution in Task~2. Scatterplots of seven of the eight covariates against the response variable $Y$ are displayed in \Cref{fig:scatter_plots}.

\subsection{Missing data}

To compare different imputation methods, we artificially created 2\% missing observations completely at random for each variable that had missing observations in the original dataset. We next imputed the missing values using four approaches: (1) imputation using the median of the available data, (2) using multivariate imputation by chained equations \citep{vanBuuren:2018} from the \textbf{R} package \texttt{mice} \citep{mice}, where the continuous variables are imputed using predictive mean matching,  (3) an iterative imputation method called missForest proposed by \cite{stekhoven2012missforest} and implemented in the \textbf{R} package \texttt{missForest} \citep{stekhoven2022missforest}, and finally (4) a generalized additive model (GAM) with smoothing splines for the continuous covariates identified previously \citep[cf.][]{Wood:2017} and implemented in the \textbf{R} package \texttt{gam}. To assess the different methods, we considered complete cases and computed the prediction root mean square error (RMSE) values for each variable using 10-fold cross-validation; results are presented in \Cref{table:missing_validation}. We observe that the \texttt{missForest} algorithm and generalized additive models outperform the median imputation and \texttt{mice} approaches in terms of prediction RMSE for each variable.  The relationship between the variables is nonlinear and complex, and thus, a random forest algorithm or GAM can better capture the dependence structure.

 Using random forests \citep{breiman2001random}, we imputed missing values by the average of numerous unpruned classification and regression trees (CARTs) for classification or regression. Through the utilization of a random forest inherent out-of-bag error estimates, \texttt{missForest} approach can estimate the imputation error without requiring a separate test set.
 The numerical implementation of the method is however markedly slower than alternatives. We did not account for the uncertainty in data imputation, and thus the precision of our estimators is probably inflated; an alternative would be multiple imputation \citep[e.g.,][]{vanBuuren:2018}, which we did not have time to explore.

\begin{table}[ht]
\centering
\caption{Prediction root mean squared error of imputed data for generalized additive models (\texttt{gam}), median imputation (\texttt{impute}), multiple imputation with predictive mean matching (\texttt{mice}), and missForest (\texttt{missForest}) methods. A smaller value indicates a better performance of the corresponding method. For Wind direction, the reported values correspond to the average Euclidean distance on the unit circle between predicted and observed angles of the holdout sets.} %
\begin{tabular}{lcccccc}
\toprule
  & ~$V_1$~ & ~$V_2$~ & ~$V_3$~ & ~$V_4$~ & ~Wind speed~ & ~Wind direction~\\
\midrule
\texttt{gam} & 3.00 & 3.78 & 5.56 & 1.19 & 1.01 & 0.71\\
\texttt{impute} & 4.23 & 5.24 & 5.77 & 1.29 & 1.11 & 1.62\\
\texttt{mice} & 4.41 & 5.73 & 7.70 & 1.50 & 0.77 & 1.35\\
\texttt{missForest} & 3.08 & 3.91 & 5.44 & 1.23 & 0.73 & 1.04\\
\bottomrule
\end{tabular}
\label{table:missing_validation}
\end{table}

\subsection{Threshold estimation}
Once the missing data are imputed, we focus on obtaining modelling extremes. The generalized Pareto distribution \citep{Davison.Smith:1990} is a theoretically justified probability model for \textit{high} threshold exceedances. Given the relationship between $Y$ and covariates in the bulk, the problem of threshold selection can be converted into a quantile regression task, specifically for choosing covariate-dependent thresholds. The $\tau$th conditional quantile is that of the conditional probability distribution of $Y$ given $\boldsymbol{X}$. In a standard quantile regression, we assume the latter is of the form $Q_{\textrm{Y}| \boldsymbol{X}}(\tau) = \boldsymbol{X}\boldsymbol{\beta}_{\tau}$. The empirical estimator is $\boldsymbol{\beta}_{\tau}$ as $\widehat{\boldsymbol{\beta}}_{\tau} = \arg \min_{\boldsymbol{\beta}_{\tau}} \sum_{i=1}^N \rho_{\tau}(Y_i - \boldsymbol{X}_i \boldsymbol{\beta}_\tau)$ where $\rho_\tau (y) = y(\tau - \mathrm{I}\{y < 0\})$ is the check function, for indicator function $\mathrm{I}\{ \cdot \}$; see \cite{koenker2005quantile} for an overview of quantile regression.

\begin{figure}[htbp!]
    \centering
\includegraphics[width = \textwidth]{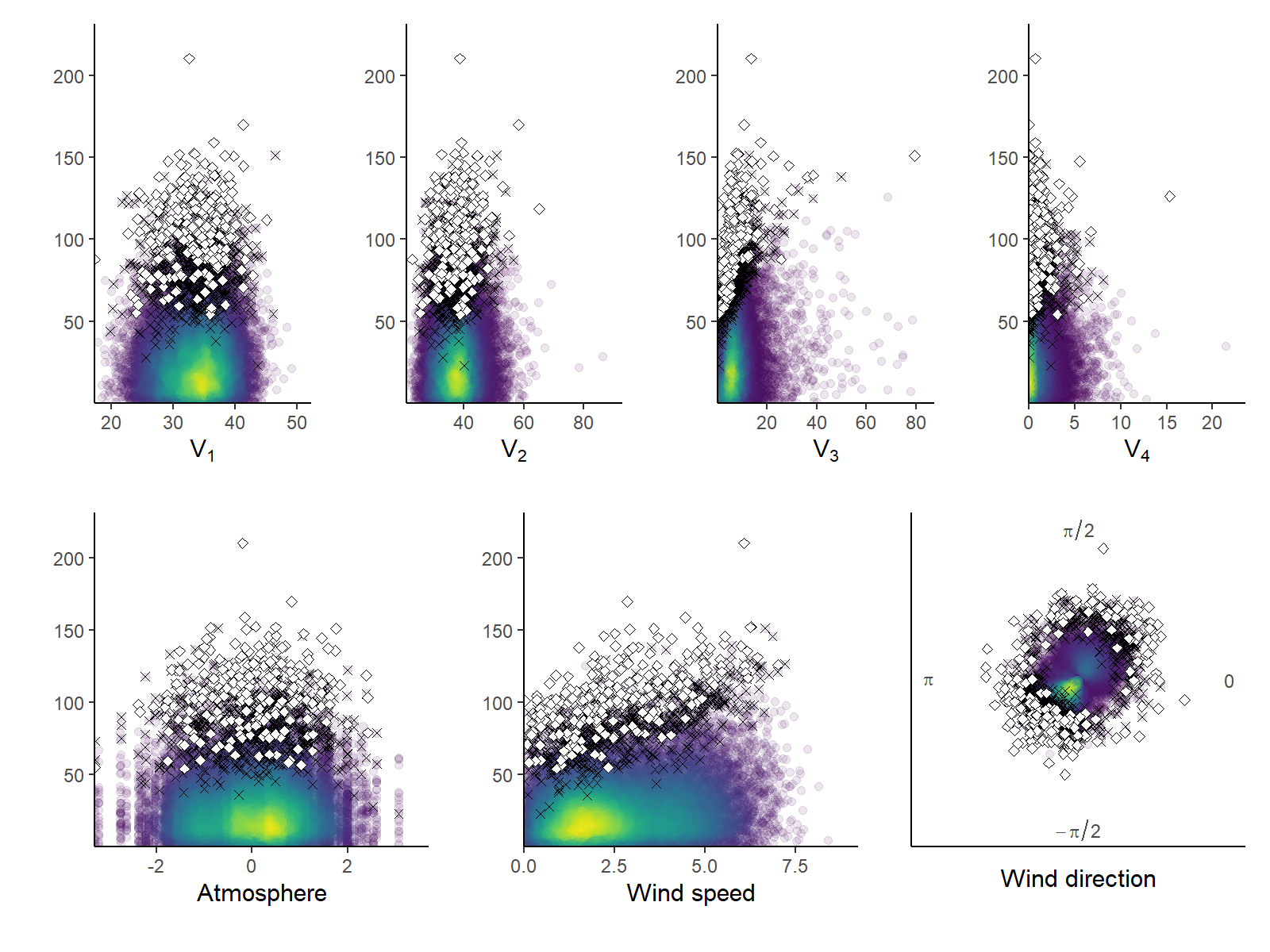}
  \caption{Bivariate kernel density and scatter plots of continuous explanatory variables against the response and scatter plots of the threshold exceedances beyond the $0.95$th (black crosses) and $0.99$th (white losange) quantile levels estimated using the asymmetric Laplace model.}
    \label{fig:scatter_plots}
\end{figure}

We also considered more flexible approaches that takes care of possible nonlinear relationships between the covariates and response variables that appear in the bivariate density estimates of \Cref{fig:scatter_plots}. There are several such statistical and machine-learning tools available in the literature. For example, the quantile regression forest approach proposed by \cite{meinshausen2006quantile}, implemented in the \textbf{R} package \texttt{quantregForest} \citep{meinshausen2017quantregForest}, performs quantile regression using random forests.

When the quantile of interest is extremely high, there are limited or no training data points surpassing it, and hence, traditional approaches for quantile regression become ineffective. To tackle this issue, \cite{gnecco2022extremal} pioneered an extremal random forest that combines generalized random forest \citep{Athey.Tibshirani.Wager:2019} for threshold estimation using a quantile loss, and uses the resulting weights to fit a generalized Pareto model to exceedances with a local likelihood whose weights arise from the random forest and with regularization of the shape over the covariate domain. This method is implemented in the \textbf{R} package \texttt{erf} \citep{gnecco2023erf}.

Another alternative is to convert the quantile regression problem into a likelihood-based model under the assumption that the observations follow the asymmetric Laplace distribution \citep{yu2001bayesian}, with density 
\begin{equation}
    f_{\textrm{ALD}}(y; \eta, \nu, \tau) = \frac{\tau (1 - \tau)}{\nu} \exp \left\lbrace -\rho_{\tau} \left( \frac{y - \eta}{\nu} \right) \right\rbrace,
\end{equation} 
where $\rho_\tau (\cdot)$ is the check function defined previously, and $\eta$, $\nu$, and $\tau$ are location, scale, and asymmetry parameters. The location parameter $\eta$, which is the $\tau$th quantile of the distribution, is used as a threshold. \cite{Youngman:2022} uses the methodology of \cite{Wood.Pya.Safken:2016} to perform generalized additive models for the parameters of asymmetric Laplace distribution with automatic selection of penalization parameters for the smooths. The location and scale parameters of the distribution are specified using cubic splines as
\begin{equation} \label{eq:evgam}
    \eta(\boldsymbol{x}) = \beta^{\eta} + \sum_{j=1}^{J_{\eta}} \sum_{k=1}^{K_j} \beta^{\eta}_{j,k} b^{\eta}_{j,k}(x_j), ~~\log \nu(\boldsymbol{x}) = \beta^{\nu} + \sum_{j=1}^{Q_{\nu}} \sum_{k=1}^{K_j} \beta^{\nu}_{j,k} b^{\nu}_{j,k}(x_j),
\end{equation}
where $\boldsymbol{x}$ denotes the covariate vector, $b^{\eta}_{j,k}$'s and $b^{\nu}_{j,k}$'s denote basis functions, and $\beta^{\eta}_{j,k_j}$'s and $\beta^{\nu}_{j,k_j}$'s denote basis function coefficients.

We compared these three approaches (\texttt{evgam}, \texttt{erf}, and \texttt{quantregForest}) numerically for the threshold selection, setting the function arguments to their default values for all three packages for a fair comparison. Using 10-fold cross-validation, we fitted the model to training data and predict the conditional quantiles at levels
 $\{0.95, 0.96, \ldots, 0.99\}$ for the holdout data. We report in \Cref{table:threshold_model_select} the percentage of responses that fall below the predicted threshold for each quantile level, accurate to 0.01. All methods appear reliable, with \texttt{erf} being closer to the target exceedance probability.

\begin{table}[ht]
\centering
\caption{Percentage of observations below the estimated thresholds for methods \texttt{quantregForest}, \texttt{erf}, and \texttt{evgam} for different quantile levels.}
\begin{tabular}{lccccc}
\toprule
  & $q_{0.95}$ & $q_{0.96}$ & $q_{0.97}$ & $q_{0.98}$ & $q_{0.99}$\\
\midrule
\texttt{quantregForest} & 94.64 & 95.51 & 96.47 & 97.50 & 98.46\\
\texttt{erf} & 94.83 & 95.80 & 96.80 & 97.92 & 99.01\\
\texttt{evgam} & 94.47 & 95.58 & 96.67 & 97.67 & 98.75\\
\bottomrule
\end{tabular}
\label{table:threshold_model_select}
\end{table}
We chose to use generalized additive models to estimate the covariate-dependent thresholds. Our model for the asymmetric Laplace distribution parameters included both categorical covariates Season and Changepoint, and smooths for all continuous covariates.
Using the \textbf{R} package \texttt{evgam}, we estimated the $0.95$th and $0.99$th quantile levels, and the threshold exceedances beyond them are presented in \Cref{fig:scatter_plots}.

\subsection{Modelling exceedances}

Once the thresholds are estimated, we need to decide on a specification for the parameters of the generalized Pareto distribution fitted to threshold exceedances; there are relatively few data points for estimation, e.g., 246 exceedances above the 99\% quantile. The scale and shape parameters, $\sigma$ and $\xi$, can be constant or modelled as smooth functions of the covariates using \texttt{evgam}. However, we found that such an approach can be numerically unstable due to the small number of exceedances, in addition to additional variance for overfitting. We considered a total of 46 generalized Pareto models with various general linear models for the log scale and shape, without smooths given the high complexity of the optimization routine and paucity of observations. For brevity, we only present seven models, whose characteristics are summarized in \Cref{table:exceedance_model_select}. Ultimately, we selected Model~2 based on the Bayesian information criterion at threshold $q_{0.98}$ and on summaries for Task~2: some of the more complex models had much lower uncertainty and seemed to overfit. We assess whether this is the case in the postmortem section.

With the fitted model, we built approximate 50\% prediction intervals for each covariate set at level 0.9999, approximating the posterior distribution of the parameters of the asymmetric Laplace distribution and the generalized Pareto models by multivariate Gaussian after integrating over the distribution of random effects of the smooths. We then took equitailed quantiles from a set of 1000 simulated posterior draws. The resulting intervals are asymmetric, and we hope they better capture the parameter uncertainty, which can be substantial for some of the more complex models at high thresholds.

\begin{table}[tbp]
\centering
\caption{Competing generalized Pareto models for threshold exceedances. The first slot is for the log scale $\log \sigma $ and the shape $\xi$. A \xmark~indicates exclusion, and \cmark~inclusion, of the covariate. }
\begin{tabular}{cccccccccc}
  \toprule
Model & $V_1$ & $V_2$ & $V_3$ & $V_4$ & WD & WS & Atmosphere & Season & Changepoint\\
  \midrule
1 & \xmark, \xmark & \xmark, \xmark & \xmark, \xmark & \xmark, \xmark & \xmark, \xmark & \xmark, \xmark & \xmark, \xmark & \xmark, \xmark & \xmark, \xmark\\
2 & \xmark, \xmark & \xmark, \xmark & \xmark, \xmark & \xmark, \xmark & \xmark, \xmark & \xmark, \xmark & \xmark, \xmark & \cmark, \cmark & \xmark, \xmark\\
3 & \xmark, \xmark & \xmark, \xmark & \xmark, \xmark & \xmark, \xmark & \xmark, \xmark & \xmark, \xmark & \xmark, \xmark & \cmark, \cmark & \cmark, \cmark \\
4 & \cmark, \xmark & \cmark, \xmark & \cmark, \xmark & \cmark, \xmark & \cmark, \xmark & \cmark, \xmark & \cmark, \xmark & \cmark, \xmark  & \cmark, \xmark\\
5 & \cmark, \xmark & \cmark, \xmark & \cmark, \xmark & \cmark, \xmark & \cmark, \xmark & \cmark, \xmark & \cmark, \xmark & \cmark, \cmark  & \cmark, \xmark\\
6 & \cmark, \xmark & \cmark, \xmark & \cmark, \xmark & \cmark, \xmark & \cmark, \xmark & \cmark, \xmark & \cmark, \xmark & \cmark, \xmark  & \cmark, \cmark\\
7 & \cmark, \xmark & \cmark, \xmark & \cmark, \xmark & \cmark, \xmark & \cmark, \xmark & \cmark, \xmark & \cmark, \xmark & \cmark, \cmark  & \cmark, \cmark\\
   \bottomrule
\end{tabular}
\label{table:exceedance_model_select}
\end{table}

Validation of conditional peaks over threshold models is inherently difficult because the data entering the likelihood depend on the covariate-dependent threshold model and the generalized Pareto only describe exceedances. Since Task~1 is judged based on coverage, we considered the interval score of \cite{Gneiting.Raftery:2007}: for a $(1-\alpha)$ equitailed interval  forecast $(l,u)$ and associated quantile response $y$, we seek to minimize the interval score
\begin{align*}
S(l, u; y) = (u-l) + \frac{2}{\alpha}(l-y)\mathrm{I}(y<l) + \frac{2}{\alpha}(y-u)\mathrm{I}(y > u).
\end{align*}
We use cross-validation to compare different functional forms for the scale and shape of a generalized Pareto model for a fixed threshold level and set of exceedances. One difficulty of the conditional specification is that the probability level for the test data to be predicted is unknown, as they depend on covariates.
 In an unconditional analysis, we could simply pick the largest $k$ points of the test sample and use their rank to infer the probability level of the holdout data, but these are functions of covariate and model dependent in our framework.

\begin{algorithm}
  \caption{Cross-validation of confidence intervals using the interval score}
\label{algo:intervalscore}
\begin{enumerate}
\item Split the $n_u$ exceedances into three folds of roughly equal size, labelled \textsf{train 1}, \textsf{train 2} and \textsf{test}.
\item Fit the generalized Pareto model separately on all data from  \textsf{train 1} and \textsf{train 2}.
\item For each tuple $(y_i, \boldsymbol{x}_i)$ from the \textsf{test} data:
\begin{enumerate}
\item  Use the estimated generalized Pareto distribution function from \textsf{train 1} to get predicted parameters for the covariate $\boldsymbol{x}_i$, say $\widehat{\sigma}_1(\boldsymbol{x}_i)$ and $\widehat{\xi}_1(\boldsymbol{x}_i)$, and obtain the probability level of the observation $y_i$, say $p_i$.
\item  Using the fitted generalized Pareto model  from \textsf{train 2}, obtain parameter estimates $\widehat{\sigma}_2(\boldsymbol{x}_i)$ and $\widehat{\xi}_2(\boldsymbol{x}_i)$, marginalizing over the smoothing parameter uncertainty, and use the latter to obtain a 50\% interval $(\widehat{l}_i, \widehat{u}_i)$ for the set of covariates $\boldsymbol{x}_i$ at probability $p_i$.
\item  Compute the observations score $S(\widehat{l}_i, \widehat{u}_i; y_i)$.
\end{enumerate}
\item Sum the scores over all observations in the \textsf{test} data.
\end{enumerate}

\end{algorithm}
With higher thresholds, it is advisable to split the data into unequal size folds and reserve more data for model fitting. Indeed, since we build our intervals using a Gaussian approximation to the sampling distribution of the vector of regression parameters $(\boldsymbol{\beta}_{\sigma}, \boldsymbol{\beta}_{\xi})$, we need the optimization algorithm to converge to ensure that the observed information matrix is positive definite at the mode. Although we did not face this issue, there is also the possibility that the shape parameter estimates are negative for the second training set (used to determine the probability level of the observation) but positive for the first training set (used for the predictions): if we get predicted probability levels of 1 when data exceeds the estimated finite upper bound, these would map to infinity.

\begin{figure}
    \centering
     \includegraphics[width = \textwidth]{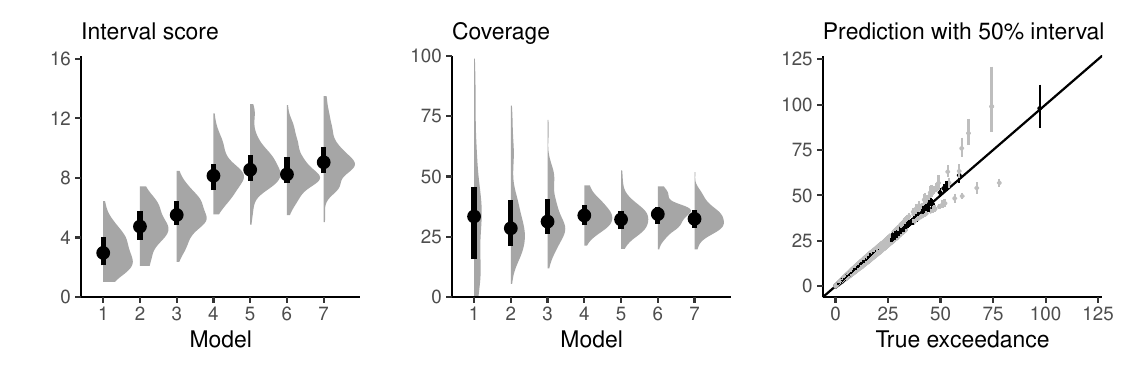}
    \caption{Density and boxplots of average interval scores (left), coverage (in percentage)  based on 100 replications of the three-fold cross-validation scheme based on exceedances above threshold $q_{0.95}$ (middle) and quantile-quantile plot of true quantiles against predictions and intervals for Model~2 based on a single repetition of the cross-validation scheme (right).}
    \label{fig:intervalscoresplots}
\end{figure}

We can repeat the cross-validation by drawing new folds at random, in order to account for the variability due to the allocation; this is illustrated in the left and middle panels of \Cref{fig:intervalscoresplots}. While the simpler Model~1 has a significantly lower interval score, the average coverage of the different models is indistinguishable even if the simpler models have more variable coverage; paired $t$-tests suggest no difference overall for coverage. The coverage for a single three-fold cross-validation for Model~2 with 50\% intervals, shown in the right panel, is 70\%.
One difficulty with generalizing this scheme is that if we were to fit thresholds in each fold to also incorporate this uncertainty, some of the data to score from the test set may be predicted to lie below the threshold.

\subsection{Performance and postmortem}

\begin{figure}[htbp!]
 \centering
\includegraphics[width = 0.9\textwidth]{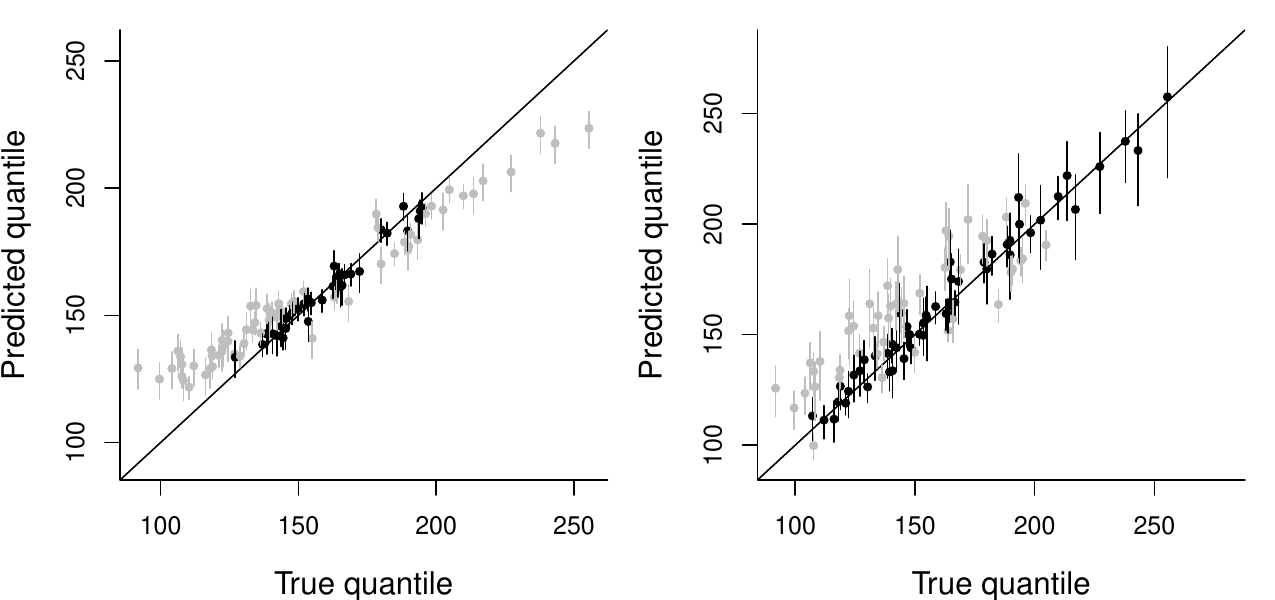}
 \caption{Quantile-quantile plot of predicted conditional quantiles for the 100 sets of covariates and 95\% intervals, against true quantiles for the final submission (left) and the best competing model (right). Intervals that cover the true quantile are shown in black, those that fail to cover in gray.}
 \label{fig:qqplot}
\end{figure}

To perform well in Task~1, we needed both accurate point predictions and good uncertainty quantification. \Cref{fig:qqplot} shows that our model shrinks considerably towards the unconditional mean due to model misspecification. In hindsight, our threshold model is overly complex and the quantile level is too high. Coupled with an overly simple generalized Pareto model, this resulted in predictions that are too high (for lower predicted values) and too low (for high predicted values), but correct on average. Of the 100 intervals, 36 covered the true values (in gray); the values were obtained using Monte Carlo simulations using the data generating mechanism described in \cite{rohrbeck2023editorial}.

Since the cross-validation scores suggested that simpler generalized Pareto models were preferable, and with seemingly comparable coverage, we opted for the simpler Model~2. We noticed that overfitting led to an important reduction in the width of the credible intervals. When calculating the intervals for the competition, we drew values from the threshold $u$, which would require in principle to refit the model. An alternative, which we consider in Task~2, is to take $u$ as fixed, but its quantile level as unknown. Due to the simplicity of our model for exceedances, the width of the credible intervals does not seem to increase with the quantile level.

To investigate the issue further, once the true data-generating mechanism was made publicly available in the editorial \citep{rohrbeck2023editorial}, we created a new test set of length 10,000 along with calculating the true 0.9999th quantiles for each combination of the explanatory variables. In \Cref{table:empirical_coverage_postmortem}, we report the proportion of cases where the estimated 50\% and 95\% credible intervals include the true extreme quantiles; a model returning the respective proportions closer to 0.5 and 0.95 is preferred. While Model 2 in \Cref{table:exceedance_model_select} with quantile level 0.98, the one we picked for the competition, provided 36\% coverage based on 50\% credible intervals, we see that the model considering all other covariates for the scale parameter of the generalized Pareto distribution returns a coverage close to 50\%. This indicates that not including the other variables ended up in over-prediction and under-prediction of the lower and upper true quantiles, respectively. Although not optimal, the model we picked in a somewhat ad hoc fashion has decent coverage when compared to other models we fitted.

\begin{table}[ht]
\centering
\caption{Empirical coverage probabilities of 50\% and 95\% credible intervals for the 0.9999th quantile of the response, based on fitting the different generalized Pareto models of \Cref{table:exceedance_model_select} at thresholds ranging from the 0.95 to the 0.99 quantiles. The first and the second entries in each table cell indicate the corresponding 50\% and 95\% credible intervals, respectively. A tuple closer to (0.5, 0.95) indicates a better predictive performance. The model selected for the competition ($\star$) and the one that would have given the best score based on the postmortem analysis ($\dagger)$ are marked in bold. Estimates are based on simulating 10\ 000 observations from the data generating mechanism and are accurate to 0.5\%.}
\begin{tabular}{crrrrr}
  \hline
Model & \multicolumn{1}{c}{$q_{0.95}$} & \multicolumn{1}{c}{$q_{0.96}$} & \multicolumn{1}{c}{$q_{0.97}$} & \multicolumn{1}{c}{$q_{0.98}$} & \multicolumn{1}{c}{$q_{0.99}$}\\
  \hline
1 & 0.140, 0.395 & 0.170, 0.470 & 0.215, 0.565 & 0.260, 0.660 & 0.070, 0.540 \\
2 & 0.175, 0.500 & 0.215, 0.590 & 0.270, 0.710 & $^\star$\scalebox{.88}[1.0]{\textbf{0.365}}, \scalebox{.88}[1.0]{\textbf{0.780}} & 0.135, 0.795 \\
3 & 0.220, 0.585 & 0.210, 0.525 & 0.240, 0.615 & 0.230, 0.745 & 0.365, 0.890 \\
4 & 0.285, 0.685 & 0.290, 0.695 & 0.320, 0.775 & 0.340, 0.815 & 0.240, 0.720 \\
5 & 0.295, 0.755 & 0.335, 0.840 & 0.405, 0.910 & $^\dagger$\scalebox{.88}[1.0]{\textbf{0.460}}, \scalebox{.88}[1.0]{\textbf{0.905}} & 0.295, 0.825 \\
6 & 0.240, 0.655 & 0.250, 0.655 & 0.285, 0.730 & 0.305, 0.765 & 0.295, 0.820 \\
7 & 0.300, 0.715 & 0.320, 0.740 & 0.355, 0.825 & 0.355, 0.880 & 0.345, 0.880 \\
   \hline
\end{tabular}
\label{table:empirical_coverage_postmortem}
\end{table}

\section{\texorpdfstring{Task~2: estimating return levels with a loss function}{Task C2}}
\label{sec:partC2}

\subsection{Data and task description}
In Task~2, teams had to provide  a point estimate of ``unconditional'' 200-year return level $\hat{q}$ that minimizes the loss function
\begin{align*}
L(q, \hat{q}) = 0.9(0.99q-\hat{q})\mathrm{I}(0.99q > \hat{q}) + 0.1(\hat{q}-1.01q)\mathrm{I}(1.01q < \hat{q}).
\end{align*}
using the `Utopia' data from Task~1.
\subsection{Loss function estimation with an unconditional model}
\begin{figure}[hbp!]
	\centering
	\includegraphics[width = \textwidth]{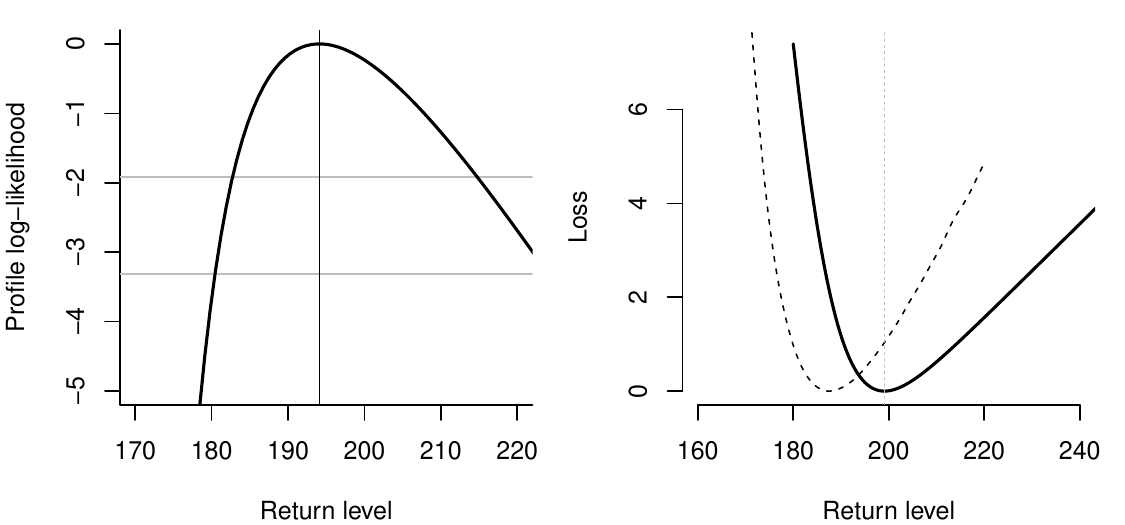}
	\caption{Profile log likelihood (left) and loss function, $0-1$ loss (dashed) and the custom loss function (right), for the unconditional 200-year return level estimated using a binomial-generalized Pareto model fitted to exceedances above the empirical 90 percentile of the original response $Y$, ignoring all covariates. The horizontal lines on the left panel indicate cutoff values for 95\% and 99\% confidence intervals. Both loss functions have been shifted so that the minimum loss is zero.}
	\label{fig:loss}
\end{figure}

To fix ideas, we consider a model that ignores all covariates: we select a constant threshold at the 90 percentile of $Y$ and fit a generalized Pareto distribution to threshold exceedances, as threshold-stability plots suggest the shape parameter is nearly constant afterward. There are $N_y=300$ observations per year and we seek a $T=200$-return level. Using the \textbf{R} package \texttt{revdbayes}, we fit a binomial-generalized Pareto model with maximal data information prior for the shape, improper prior for the log scale, and beta prior for the probability of exceedance. We draw 10\ 000 independent samples from the corresponding posterior, with  $\boldsymbol{\theta} = (\sigma_u, \xi, \zeta_u)$ the vector of parameters consisting of the scale, shape, and probability of exceedance above $u$. The $T$-years return level is $q(\boldsymbol{\theta})=u + \sigma_u/\xi\{(N_yT\zeta_u)^\xi-1\}$. We approximate the loss function $L$ pointwise by averaging the loss over the posterior samples to obtain the return level estimate as
\begin{align*}
	\hat{q}^{\star} = \operatornamewithlimits{argmin}\limits_{\hat{q}}\int_{\boldsymbol{\Theta}} L\{q(\boldsymbol{\theta}), \hat{q}\} p(\boldsymbol{\theta} \mid \boldsymbol{y})\mathrm{d} \boldsymbol{\theta}.
\end{align*}
The right-hand panel of \Cref{fig:loss} shows the loss function provided by the organizers, which is minimized by taking a return level of 198. Also displayed is the 0-1 loss for the posterior, which yields a lower point estimate. Virtually similar inferences are obtained using the inhomogeneous Poisson point process formulation.
\subsection{Averaging for covariate-dependent extreme value models}
Consider next extreme value models in which the threshold and generalized Pareto model parameters may depend on covariates; we use the same fitted models as in Task~1. We have access to a random sample of  $n$ independent and identically distributed responses $Y_1, \ldots, Y_n$ with associated covariates vectors $\boldsymbol{x}_1, \ldots, \boldsymbol{x}_n$ and parameter vector $\boldsymbol{\theta}$. The unconditional distribution of $Y$ is, by the law of total probability,
\begin{align*}
\Pr(Y \leq z) = \int_{\mathcal{X}} \Pr\{Y(\boldsymbol{x}) \leq z \mid \boldsymbol{X}=\boldsymbol{x}\} \Pr(\boldsymbol{X}=\boldsymbol{x})\mathrm{d} \boldsymbol{x}.
\end{align*}

We consider sampling with replacement $B$ observations from the set of covariates $\{\boldsymbol{x}_i\}_{i=1}^n$ for each posterior draw $\boldsymbol{\theta}$ of the parameter vector.
A $T$-year return level is, by definition, the level exceeded by an annual maximum with probability $1/T$. With $m$ observations per year, the $(1-1/T)$th unconditional quantile $q(\boldsymbol{\theta})$ is the value $z$  that solves
 \citep[][\S~2.3]{Youngman:2022}
\begin{align}
\left[\frac{1}{B} \sum_{i=1}^B F\{z; u(\boldsymbol{x}_i, \boldsymbol{\theta}), \sigma(\boldsymbol{x}_i, \boldsymbol{\theta}), \xi(\boldsymbol{x}_i, \boldsymbol{\theta}), \zeta_u(\boldsymbol{x}_i, \boldsymbol{\theta})\}\right]^m = 1-1/T,\label{eq:retlev}
\end{align}
where $F(\cdot)$ is the estimated distribution function of the binomial-generalized Pareto,
\begin{align*}
 F(z; u, \sigma, \xi, \zeta_u) = 1-\zeta_u  \left(1+\xi \frac{z - u}{\sigma}\right)_{+}^{-1/\xi}.
\end{align*}
We can take logarithms on both sides of the return level equation and use a root-finding algorithm to obtain $z$. We use the \texttt{evgam} package for inference \citep{Youngman:2022}. The posterior draws are obtained from a Gaussian approximation to the regression coefficients, incorporating the random effect uncertainty for smooths.

To account for the uncertainty arising from the covariate distribution, we could use a nonparametric bootstrap  and sample observations with replacement from complete cases. In the Bayesian paradigm, the equivalent would be the Bayesian bootstrap \citep{Rubin:1981}, i.e., drawing the vector of probability from a Dirichlet vector with weight vector $\boldsymbol{\alpha}=\mathbf{1}$. We could also sample new data for $(V_1, V_2), (V_3, \text{Season}), V_4$, etc., separately for each group of covariates identified in the exploratory data analysis. Then, for each vector of posterior draws $\boldsymbol{\theta}_i \sim p(\boldsymbol{\theta} \mid \boldsymbol{X})$, we compute the return level as $q(\boldsymbol{\theta}_i; \widetilde{X})$ and repeat this procedure to get a posterior sample of return levels. Note that this approach differs from posterior predictive inference \citep[][\S~2.3]{Northrop:2017}.

\Cref{tbl:retlevs} gives the estimated return levels after averaging over the distribution of covariates by resampling rows from the complete cases with replacement; this amounts to using a nonparametric bootstrap. We take $u$ as fixed, but the probability of exceedance $\zeta_u(\boldsymbol{x})$ as unknown: the parameters of the binomial-generalized Pareto model $\zeta_u(\boldsymbol{x})$ and $\{\sigma(\boldsymbol{x}), \xi(\boldsymbol{x})\}$ are orthogonal and so we simply draw from the posterior of the asymmetric Laplace and generalized Pareto distributions separately to reflect the global uncertainty.  Similar results are obtained for the Bayesian bootstrap (not shown), where we resample columns independently by block; these lead to a 3\% increase in the average return levels.

\begin{table}[htbp!]
\centering
\caption{Unconditional return level (standard errors) for different generalized Pareto models as a function of the threshold. Models components are given in \Cref{table:exceedance_model_select}.}
 \label{tbl:retlevs}

\begin{tabular}{crrrr}
\toprule
 Model & \multicolumn{1}{c}{$q_{0.95}$} &\multicolumn{1}{c}{$q_{0.96}$}& \multicolumn{1}{c}{$q_{0.97}$} & \multicolumn{1}{c}{$q_{0.98}$}\\
\midrule
1 & 183.0 (0.03) & 182.2 (0.02) & 187.7 (0.04) & 195.5 (0.07)\\
2 & 181.5 (0.03) & 182.5 (0.02) & 189.9 (0.05) & 202.0 (0.10)\\
3 & 189.7 (0.05) & 195.5 (0.06) & 213.8 (0.14) & 239.6 (0.20)\\
4 & 209.5 (0.11) & 200.4 (0.05) & 202.5 (0.06) & 209.6 (0.08)\\
5 & 226.0 (0.23) & 210.8 (0.10) & 207.7 (0.10) & 213.8 (0.14)\\
6 & 215.0 (0.15) & 210.7 (0.06) & 217.8 (0.09) & 234.4 (0.17)\\
7 & 225.3 (0.19) & 215.4 (0.10) & 219.0 (0.10) & 237.7 (0.21)\\
\bottomrule
\end{tabular}
\end{table}

In the initial submission, we did not have time to consider Task~2 and submitted the maximum likelihood estimate of the return level (189 units). Given our dismal ranking, we gave more prior weight to models that returned higher return levels: as the loss function penalizes smaller values, our guess was that we had underestimated the true quantile. This in turn favored models that had risk estimates close to the naive unconditional estimator without covariate. We picked a single model estimated models using the imputed value from Task~1 and used the latter to average over the covariate distribution by giving equal weight to each of the 21 000 observations, including the imputed values. We treated the threshold as random and the probability of exceedance $\zeta_u$ as a fixed quantity. This gave us a value of 201.84 units for the quantile level.

\subsection{Postmortem}
Owing to the lack of time, we cut corners and used all observations (including imputations) for the prediction to compute return levels, rather than resampling with replacement. We also fixed the probability of exceedance  $\zeta_u$ in the submission and varied the thresholds $u$ instead.

After the competition, we tried taking multiple imputations for the missing data and taking the average threshold level for each of these. While the threshold levels were strongly correlated with the single imputations, this led to noticeable differences in the results for Task~2, with much higher unconditional return levels. This suggests that the results are quite sensitive to these imputed values when extrapolating at extreme levels.

We conjecture that very high values of the return levels are an artefact of the threshold overfitting, which leads to abnormally high values for $u(\boldsymbol{x})$. Since the distribution function in \Cref{eq:retlev} reduces to $1-\zeta_u$ below the threshold, irrespective of the value of $\hat{q}$, this shifts the loss function to the right. A semiparametric model would assign a much lower probability level than the binomial-generalized Pareto model.

One important aspect that we knowingly ignored is model uncertainty: we picked the model from Task~1 based on pragmatic considerations, and some of the models that would have done better for coverage would have led to overestimation. To combine multiple models, we could assign each of them prior weights and perform Bayesian model averaging \citep[cf.][]{Raftery:1995}.
One could use pseudo Bayesian averaging by working with cross-validation predictive densities, obtained through importance sampling as in \cite{Northrop:2017}.

\section{\texorpdfstring{Task~3: trivariate problem with mixed dependence}{Task C3}}
\label{sec:partC3}
\subsection{Data and task description}
The data set for Task~3 contains 21\ 000 observations representing ``70 years of daily time series'' for a trivariate series with standard Gumbel margins, along with two covariates (Season and Atmosphere). The goal of Task~3 is to estimate the joint probability of extreme events at all three sites.  Specifically, if $Y_1$, $Y_2$ and $Y_3$ denote the three variables on the standard Gumbel scale, we estimate
\begin{align}
p_1 &= \Pr(Y_1>y, Y_2>y, Y_3>y), \nonumber \\
p_2 &= \Pr(Y_1>v, Y_2>v, Y_3<m),
\label{eq:ProbC3}
\end{align}
where $y=6$, $v=7$, and $m=-\log\{-\log(2)\}$ is the median of  a standard Gumbel variate.

\subsection{Measuring and modelling multivariate tail dependence}
Most extreme value methods can only characterize events away from the origin when all variables are simultaneously large. The probability $p_1$ in \cref{eq:ProbC3} is an example of such, since the region of interest $(y,\infty)^3$ lies along the diagonal in the positive orthant. For $p_2$, the risk region is $(v,\infty)^2\times (-\infty,m)$, but we can filter to keep only data for which $Y_3$ is below the median, so that
$p_2 = 0.5\Pr(Y_1>v, Y_2>v \mid Y_3 < m)$.

We considered three different techniques to estimate the multivariate extremes probability: (1) a semiparametric model exploiting the hidden regular variation framework \citep{ledford1997modelling}; (2) the conditional extreme value model of \cite{Heffernan.Tawn:2004}, and (3) the semiparametric model exploiting the geometric approach of \cite{Wadsworth.Campbell:2022}. We did not revisit this task between the initial and final submission and did not consider covariates at all, as we did not find an obvious pattern for the dependence in our exploratory data analysis.

\subsubsection{Tail dependence measures} \label{sec:taildep}
Consider a random vector $\boldsymbol{Y}$ with known marginal distribution functions $F_i (i=1, \ldots, D)$.
Two summaries commonly employed to describe the joint tail behaviour of $\boldsymbol{Y}$ are the tail correlation $\chi$ and the coefficient of tail dependence $\eta$.
The tail correlation coefficient at level $v$ is
\begin{align*}
\chi(v) =\frac{\Pr\{F_1(Y_1) > v, \ldots, F_D(Y_D)>v\}}{1-v}, \qquad v \in (0, 1).
\end{align*}
Since the marginal distributions are continuous, \(\sum_{i=1}^n \mathrm{I}(\min_i \{F_i(X_i)\} >v) \sim \mathsf{Bin}(n, p_v)\) and a point estimator of the tail correlation is
\(\widehat{\chi}(v) = \widehat{p}_v/(1-v)\) with associated variance \(n^{-1}\widehat{p}_v(1-\widehat{p}_v)/(1-v)^2\), where $\widehat{p}_v$ is the maximum likelihood estimator of the probability of exceedance.
If $\lim_{v \to 1} \chi(v)=0$, we say that the vector exhibits asymptotic independence, and asymptotic dependence otherwise.

In the case of asymptotic independence, $\chi$ is not a useful descriptor of the strength of dependence and tells us nothing about the rate of decay of the joint tail. If we map data to standard Pareto margins and compute the structural variable $T_{\mathrm{p}}=\min_{j=1}^D \{1-F_i(Y_i)\}^{-1}$, we can define $\eta \in (0,1]$ implicitly through the relation \citep[][Section 5]{Ledford:1996}
\begin{align}
\Pr(T_{\mathrm{p}} >x) = \mathcal{L}(x) x^{-1/\eta}, \label{eqHRV}
\end{align}
where $\mathcal{L}(x)$ is a slowly-varying function, i.e., $\mathcal{L}(cx)/\mathcal{L}(x) \to 1$ as $x\to \infty$. The variables are positively associated if $\eta \in (1/D, 1]$, independent if $\eta = 1/D$ and negatively dependent otherwise. In the case of asymptotic dependence, $\eta=1$. \citet[][\S~4]{deHaan.Zhou:2011} details properties of the tail dependence coefficients.

\begin{figure}[htbp!]
    \centering
    \includegraphics[width = 0.49\textwidth,valign=t]{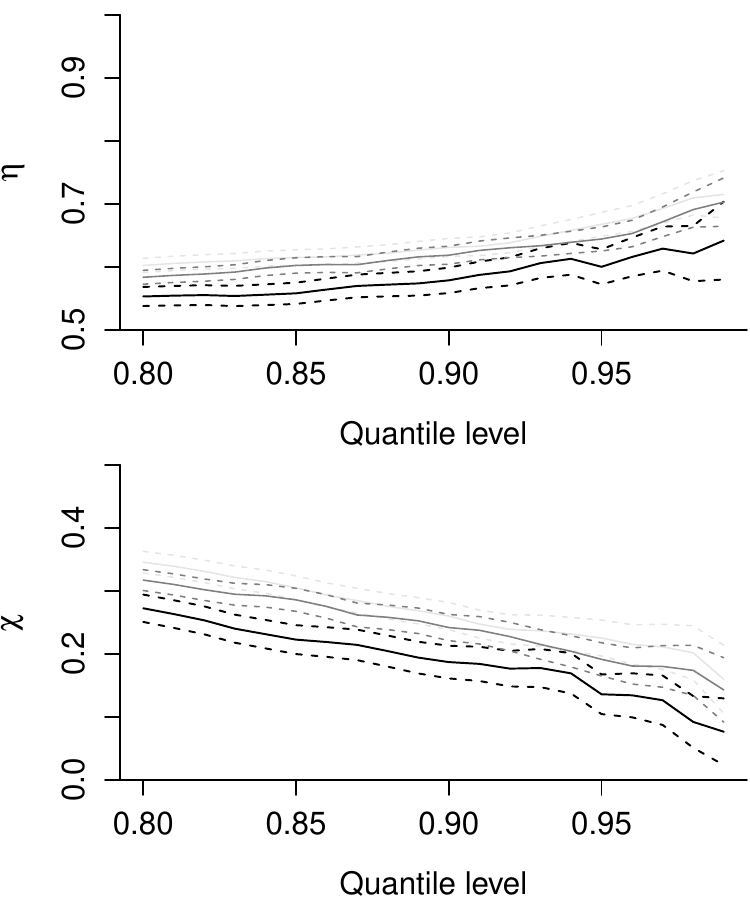}
    \includegraphics[width = 0.5\textwidth,valign=t]{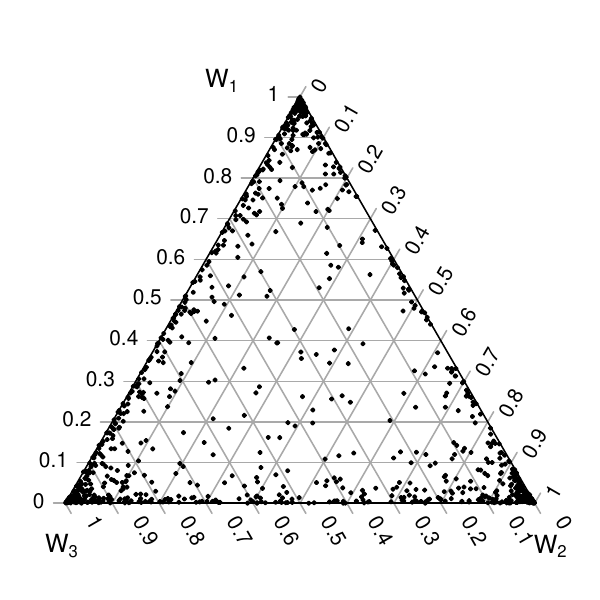}
    \caption{Left: Pairwise coefficients of tail dependence $\eta$ (top) and tail correlation $\chi$ (bottom) for pairs $\{Y_1, Y_2\}$ (black), $\{Y_1, Y_3\}$ (light gray) and $\{Y_2, Y_3\}$ (dark gray). Right: scatterplots of angles $\boldsymbol{W}=\boldsymbol{F}/\|\boldsymbol{F}\|_1$ on unit Fréchet margins whose radial component $\|\boldsymbol{F}\|_1$ exceeds its 0.95 quantile. }
    \label{fig:chiplots-C3}
\end{figure}

For the data of Task~3, all pairs seem to exhibit some degree of dependence, but do not seem to show asymptotic dependence (as estimated values of $\eta$ are far from one), while the estimates of the tail correlation $\widehat{\chi}$ decrease towards zero as the threshold increases; see the left panels of \Cref{fig:chiplots-C3}. We also produced plots of $\chi$ and $\eta$ for each pair, splitting the data by Season, but found no visible difference.

If we further consider a pseudo-polar decomposition of the data after mapping margins to the unit Fréchet scale, we find that there is strong visual evidence of asymptotic independence with mass on the vertices and edges of the simplex, as shown in the right panel of \Cref{fig:chiplots-C3}.

The assumption of hidden regular variation of \cref{eqHRV} allows one to extrapolate the probability of rare sets beyond the range of the data: if we map observations to unit exponential scale with $E_i = -\log \{1- F(Y_i)\}$ and compute  $T_{\mathrm{e}} = \min_{j=1}^D E_{j}$, then we have the asymptotic approximation
\begin{align*}
 \Pr(T_{\mathrm{e}}  \in u+t +A \mid T_{\mathrm{e}} \in u + A) \sim \exp(-t/\eta),
\end{align*}
for a set $u+A$ lying above the high threshold $u$. Treating exceedances above $u$ as an exponential sample and setting $A = (0, \infty)^D$, the maximum likelihood estimator of the scale parameter $\eta$ is the sample mean of the structural variable $T_{\mathrm{e}}$, truncated above at 1.

\subsubsection{Conditional extremes approach}
\label{sec:heffernantawn}
Although $\eta$ and $\chi$ provide summaries of extremal dependence, they do not fully characterize the tail. An alternative approach to estimate the two probabilities in \Cref{eq:ProbC3} is to examine the conditional joint distribution of $D-1$ variables conditional on exceedance of $D$th variable. With $\boldsymbol{L}$ denoting the $D$ vector in standard Laplace margins, the conditional extreme value model of \cite{Heffernan.Tawn:2004} assumes that, for large $u$, the probability distribution of $\boldsymbol{L}_{-j}$ given exceedance of $L_j$ can be approximated by
\begin{align*}
\Pr\left(\left.L_j^{-\boldsymbol{\beta}}({\boldsymbol{L}_{-j} - \boldsymbol{\alpha} L_j}) \in \cdot, L_j> u + t \,\,\right|\, L_j>u \right) \approx \Pr(\boldsymbol{Z} \in \cdot) \exp(-t), \qquad u \to \infty.
\end{align*}
To estimate model parameters, we assume that the residual vector $\boldsymbol{Z}$ follows a $D-1$ multivariate Gaussian with mean $\boldsymbol{\mu}$ and covariance matrix $\diag(\boldsymbol{\sigma}^2)$. After estimating the $4(D-1)$ model parameters $\boldsymbol{\alpha} \in [-1,1]^{D-1}$ and $\boldsymbol{\beta}\in (-\infty,1)^{D-1}$, $\boldsymbol{\mu} \in \mathbb{R}^{D-1}$ and $\boldsymbol{\sigma} \in \mathbb{R}_{+}^{D-1}$, we can obtain the empirical residuals
\begin{align}
\widetilde{\boldsymbol{z}}_i = \frac{\boldsymbol{l}_{i,-j} - \hat{\boldsymbol{\alpha}} l_{i,j}}{l_{i,j}^{\hat{\boldsymbol{\beta}}}}, \qquad i=1, \ldots, n. \label{eq:resid}
\end{align}
\cite{Heffernan.Tawn:2004} propose to estimate the joint tail probabilities of events falling in a risk region which is a subset of $L_j > v$ for $v \geq u$ by first simulating $L^*_j - v \sim \mathsf{Exp}(1)$, then drawing a residual vector $\boldsymbol{Z}^*$ with replacement from the empirical distribution of \cref{eq:resid} and setting $\boldsymbol{L}^{*}_{-j} = \hat{\boldsymbol{\alpha}} L^{*}_j + L_j^{*\hat{\boldsymbol{\beta}}} \boldsymbol{Z}^*$. The probability of interest is estimated by calculating the proportion of simulated points falling in the risk region, times the probability of exceedance of the conditioning variable.

\subsubsection{Geometric approach}

An alternative methodology involves geometric extremes \citep{Nolde.Wadsworth:2022}. With data in standard exponential margins, we consider the scaled cloud of points $\{E_i/\log n\}$ $(i=1, \ldots, n)$ and assume the latter converges onto the limit set $G= \{\boldsymbol{x} \in \mathbb{R}_{+}^3: g(x) \leq 1\}$. The limit set is characterized by the gauge function $g$, a one-homogeneous function which fully characterizes the multivariate asymptotic dependence structure.
We describe succinctly the methodology of \cite{Wadsworth.Campbell:2022}, which was used to estimate $p_1$ and $p_2$. First, we consider a radial-angular decomposition of the standardized exponential variates, $R=\sum_{j=1}^3E_j$ and $\boldsymbol{W}=\boldsymbol{E}/R$.
We then use sliding windows over different angles to obtain a high radial threshold $r_0(\boldsymbol{w})$ at a fixed quantile level $1-\alpha$ and extract exceedances. For a parametric gauge function $g(\cdot)$, we fit the model via maximum likelihood assuming
\begin{align}
\label{eq:trucGamma-geomExt}
R \mid \{\boldsymbol{W} = \boldsymbol{w}, R > r_0(\boldsymbol{w})\} \sim \textsf{truncated gamma}\{\alpha, g(\boldsymbol{w})\},
\end{align}
where $\alpha$ and $g(\boldsymbol{w})$ are the shape and rate parameters, respectively, of the gamma distribution truncated above $r_0(\boldsymbol{w})$.
 We took the gauge function of the Gaussian distribution with covariance matrix $\boldsymbol{\Sigma}$,  $g(\boldsymbol{w})=\boldsymbol{w}^{1/2\top}\boldsymbol{\Sigma}^{-1}\boldsymbol{w}^{1/2}$, where the square root denotes a componentwise-operation, as model for the dependence.

Inference for extreme levels is performed via Monte Carlo methods. Specifically, let $R' = R/r_0(\boldsymbol{w})$; the probability of observations falling in set $B$ can be calculated using the relationship
\begin{align*}
\Pr(\boldsymbol{Y} \in B) = \Pr(\boldsymbol{Y} \in B \mid R'>1) \Pr(R'>1).
\end{align*}
We first sample $\boldsymbol{W} \mid R'>1$ from the empirical distribution of angles, then simulate conditional on that draw from the fitted truncated gamma distribution of $R \mid \{\boldsymbol{W} = \boldsymbol{w}, R > r_0(\boldsymbol{w})\}$ in \cref{eq:trucGamma-geomExt}. The term $\Pr(R'>1)$ can be estimated using the proportions of points $R'$ exceeding 1. We fitted the model using the \textbf{R} package \texttt{geometricMVE} and report results in \Cref{tab:comp-C3} for four different quantile levels of the radial threshold, along with those for the conditional extremes and \cite{Ledford:1996} approach.

\begin{table}[htbp]
  \centering
  \caption{Probability estimates $(\times 10^6)$ for $p_1$ and $p_2$ based on the Heffernan--Tawn conditional extremes model (conditional), hidden regular variation (HRV), and the geometric extremes approach (geometric) for thresholds at different quantile levels. Monte Carlo estimates are accurate to $10^{-8}$, i.e., to two significant digits. }
  \label{tab:my-table}
  \begin{tabular}{lrrrrrrrrrr}
\toprule
 & \multicolumn{5}{c}{$p_1$} & \multicolumn{5}{c}{$p_2$} \\
\cmidrule(lr){2-6}  \cmidrule(lr){7-11}
 method & \multicolumn{1}{c}{$q_{0.90}$}  & \multicolumn{1}{c}{$q_{0.95}$} & \multicolumn{1}{c}{$q_{0.96}$} & \multicolumn{1}{c}{$q_{0.97}$} & \multicolumn{1}{c}{$q_{0.98}$} & \multicolumn{1}{c}{$q_{0.90}$}  & \multicolumn{1}{c}{$q_{0.95}$} & \multicolumn{1}{c}{$q_{0.96}$} & \multicolumn{1}{c}{$q_{0.97}$} & \multicolumn{1}{c}{$q_{0.98}$} \\ \midrule
conditional & 19.12 & 26.39 & 24.16 & 28.23 & 22.43 & 3.66 & 8.53 & 7.53 & 14.14 & 5.29\\
HRV & 3.48 & 5.00 & 5.12 & 5.30 & 6.28 & 2.52 & 5.56 & 7.44 & 8.85 & 11.40\\
geometric & 4.30 & 5.16 & 5.37 & 6.10 & 6.40 & 1.82 & 3.41 & 3.14 & 5.37 & 5.69\\
\bottomrule
\end{tabular}
   \label{tab:comp-C3}
\end{table}

\subsection{Postmortem}
In our final submission, we reported the estimates obtained by fitting the Heffernan--Tawn model with both the marginal and dependent thresholds set to $0.95$ quantile each using the \texttt{texmex} package. Our probability estimates were $1.56 \times 10^{-5}$ for $p_1$ and $6.80 \times 10^{-6}$ for $p_2$; the discrepancy with \Cref{tab:comp-C3} is due to the (unnecessary) estimation of the marginal distributions by \texttt{texmex}.  Since we were first in the initial ranking and were short on time, we did not revisit the task. A simple way of including the covariates would have been to let the parameters of the conditional extremes model vary as in \cite{Jonathan.Ewans.Randell:2013}, or by doing the semiparametric extrapolation separately for each Season.

\section{\texorpdfstring{Task~4: predicting the probability of simultaneous exceedance in high-dimensional multivariate model}{Task C4}}
\label{sec:partC4}
\subsection{Data and task description}
The data for task \(4\) consists of 10 000 observations from a 50-dimensional random vector with standard Gumbel margins. The variables are split in two equal-sized sets $U_1$ and $U_2$: we seek to estimate $p_1$, the joint probability of exceedance of all variables beyond the marginal quantile at level  $\phi_1=1/300$ for variables in $U_1$ and $\phi_2=12/300$ for variables in $U_2$. The second target, $p_2$, is the joint probability of exceedances of all variables beyond $F^{-1}(1-\phi_1)$.
\subsection{Exploratory data analysis}
\begin{figure}[htbp!]
\centering
\includegraphics[width=0.9\textwidth]{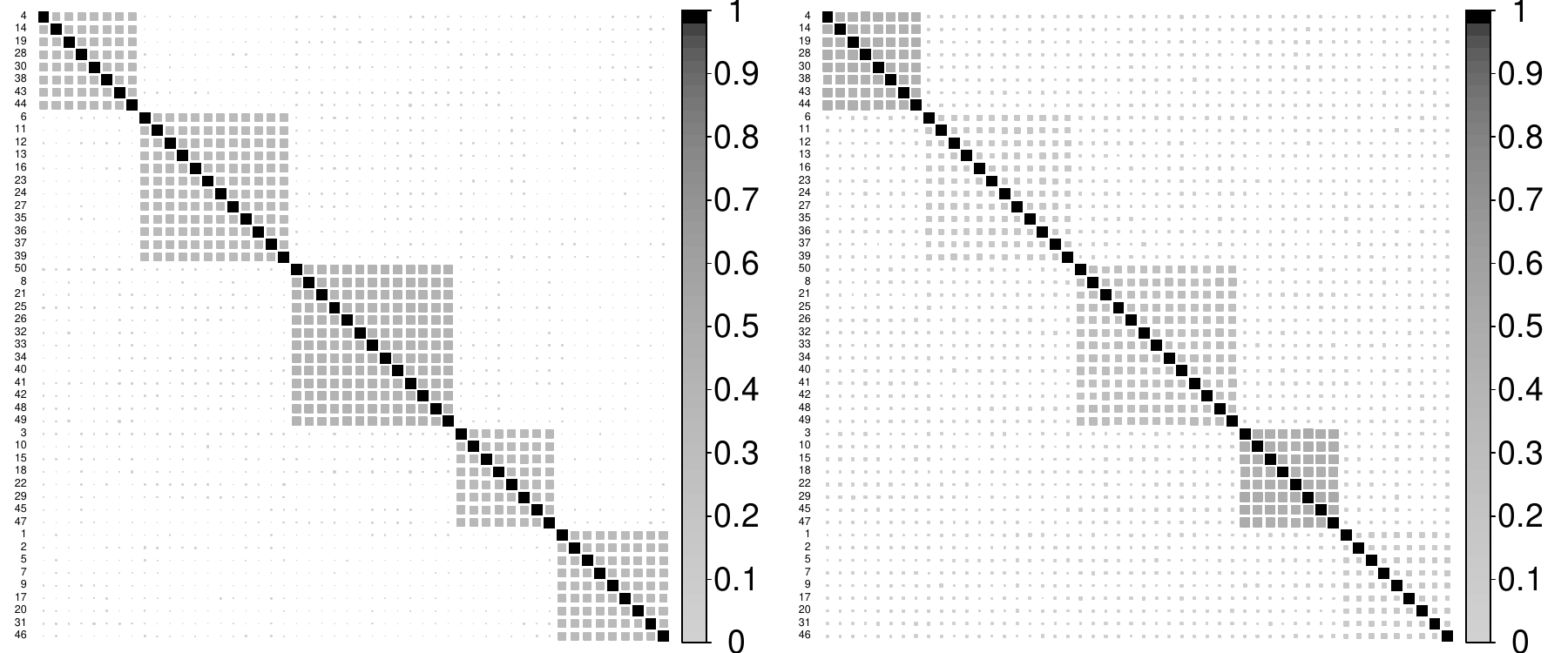}
\caption{Matrices of Kendall's \(\tau\)
correlations
(left) and pairwise tail correlation coefficient \(\chi(0.975)\)
(right). Clusters have been identified using hierarchical clustering and reordered to match the order of the ones described in the data challenge editorial.}
\label{fig-C4corrplots}
\end{figure}

Since this is a fairly high-dimensional multivariate problem, it is helpful to
investigate the dependence structure first to try to break the problem into smaller components. The left-hand panel of
\Cref{fig-C4corrplots} shows the estimated Kendall's $\tau$
correlation matrix after permuting stations accordingly to clusters estimated using hierarchical clustering with Ward's method. The correlation matrix of \Cref{fig-C4corrplots} suggests a block structure with a compound
symmetry structure within a cluster, with a within-block correlation ranging between 0.3
and 0.45.

\subsubsection{Testing for partial
exchangeability}\label{testing-for-partial-exchangeability}
We tested for partial exchangeability based on Kendall's $\tau$
matrix \(\mathbf{T}_n\) using results from \cite{Perreault:2022}, who study
the asymptotic behaviour of the
\(p=n(n-1)/2\) vector obtained by stacking columns of the upper triangle
of \(\mathbf{T}_n\), \(\widehat{\boldsymbol{\tau}}_{p}\). The test
looks at the difference between \(\widehat{\boldsymbol{\tau}}_{p}\) and
the constrained version obtained by projecting the group structure using
a \(p \times L\) projection matrix \(\mathbf{B}\) enforcing the cluster
structure. We form the orthogonal projection matrix
\(\mathbf{P} = \mathbf{I}_p - \mathbf{B}\mathbf{B}^{\leftarrow}\) for
the difference where \(\mathbf{B}^{\leftarrow}\) is the Moore--Penrose
generalized inverse of \(\mathbf{B}\). In our setting, the constrained
model has \(L=6\) different entries for the 1250 estimates,
corresponding to the pairwise entries of the five different clusters;
all other pairs are pooled in a single entity.

We considered two statistics,
\(E_{n} = \|(n\widehat{\boldsymbol{\Sigma}}_{np})^{-1/2}\mathbf{P}\widehat{\boldsymbol{\tau}}_{p}\|_2\)
and
\(M_n = \|(n\widehat{\boldsymbol{\Sigma}}_{np})^{-1}\mathbf{P}\widehat{\boldsymbol{\tau}}_{p}\|_{\infty}\),
where \(\widehat{\boldsymbol{\Sigma}}_{np}\) is the jackknife estimator
of the covariance matrix of \(\boldsymbol{\tau}_{p}\) obtained by
averaging entries to enforce the postulated sparsity structure \citep{Perreault:2019}. According to
 Propositions 5.1 and 5.2 (b) of \cite{Perreault:2022}, the asymptotic null distribution of the statistics coincides with
that of \(\|\boldsymbol{Z}\|_2\) and \(\|\boldsymbol{Z}\|_{\infty}\),
where
\begin{align*}
\boldsymbol{Z} \sim \mathsf{Normal}_p(\boldsymbol{0}_p, \boldsymbol{\Sigma}_{p}^{-1/2}\mathbf{P}\boldsymbol{\Sigma}_{p}\mathbf{P}\boldsymbol{\Sigma}_{p}^{-1/2}).
\end{align*}
We replace the unknown \(\boldsymbol{\Sigma}_p\) by the estimated matrix
\(n\widehat{\boldsymbol{\Sigma}}_{np}\). Monte Carlo estimates of the
\(p\)-values are 0.74 for \(E_n\) and 0.64 for \(M_n\), suggesting no
evidence against the null of partial exchangeability. The asymptotic
null distribution for \(E_n\) is  \(\chi^2_{p-L}\) and both Monte Carlo and asymptotic $p$-value estimates are nearly identical.

\subsubsection{Extremal dependence}

\begin{figure}
\centering
\includegraphics[width = 0.9\textwidth]{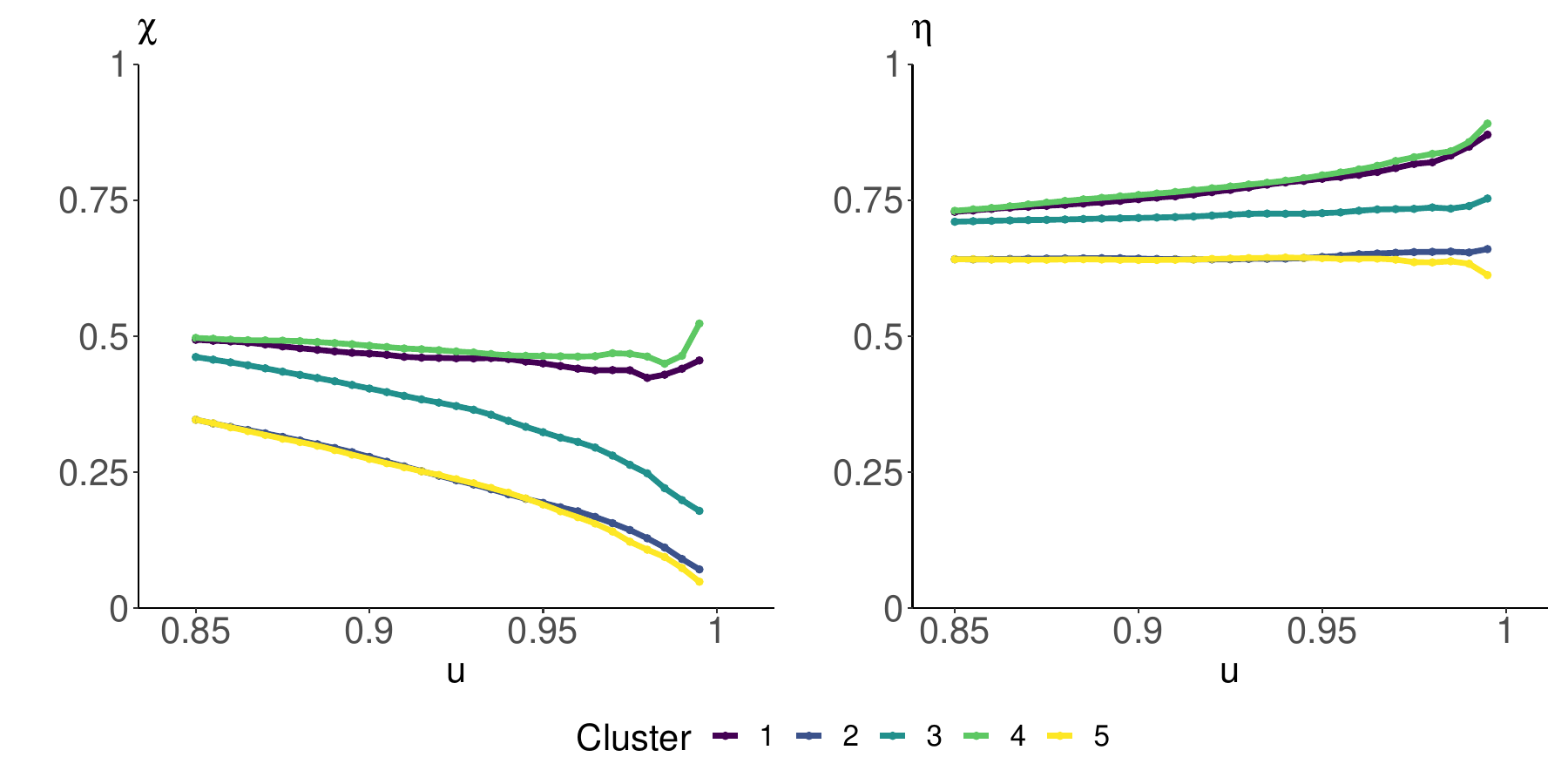}
\caption{Average pairwise estimates of the  tail correlation $\chi$ (left) and empirical tail dependence $\eta$ (right) per cluster, obtained by pooling estimates over all pairs, as a function of the quantile level $u$.}
\label{fig-taildep}
\end{figure}

We used the tail dependence coefficient introduced in \Cref{sec:taildep} to assess the degree of extremal dependence; these estimates, plotted in the right panel of \Cref{fig-C4corrplots} again suggest a lack of dependence in the tail for stations in different blocks and lack of any asymptotic dependence between any pair from a different cluster. Under the assumption of exchangeability, we produce plots of the tail correlation \(\chi(v)\) and of the
coefficient of tail dependence \(\eta(v)\) obtained using empirical
estimators at probability level \(v \in \{0.8, \ldots, 0.99\}\) for all pairs from the identified clusters.
These pooled pairwise estimates, shown in \Cref{fig-taildep},
reveal the following patterns:

\begin{itemize}
\item
  Clusters 2, 3 and 5: stable estimates of \(\eta(v)\) but far from unity,
  \(\chi(v)\) decreasing towards zero as $v \to 1$: both indicators are suggestive of asymptotic independence
  \citep{Coles.Heffernan.Tawn:1999}.
\item
  Clusters 1 and 4: \(\eta(v)\)  nearly constant or increasing towards 1, with \(\chi(v)\) more or
  less constant at \(0.5\), somewhat coherent with asymptotic dependence.
\end{itemize}

As a dimensionality-reduction step, we assume hereafter that
observations from different clusters are independent; this reduces the
problem to estimating separately the joint probability of exceedances of
\(\Pr(Y_{i_j} > s_i; i_j \in C_k)\) for each cluster as, under independence, the quantity of
interest is the product of the probabilities in each cluster.

\subsection{Semiparametric extrapolation}
As a preliminary approach, we considered semiparametric estimation for $p_2$  using the approach of  \citet[][Section 5]{Ledford:1996}; see \Cref{sec:taildep} for an overview. For each cluster, we constructed the structural variable $T_{\mathrm{e},k} = \min_{i \in C_k} E_i$, where $E_i$ again denotes the random variables on standard exponential margins. We computed the probability of joint exceedance in each cluster by setting the threshold $u$ equal to the marginal 98.5\% empirical percentile of $T_{\mathrm{e},k}$ and set $u + t = -\log(\phi_1)$.

One drawback of the Ledford--Tawn approach is that all components must decay at the same rate. \cite{Wadsworth.Tawn:2013} consider extrapolation along rays from the origin in different directions, leading to the approximation
\begin{align*}
 \lim_{u_k \to \infty}\Pr(E_i > \beta_i (t_k+u_k) \mid E_i > \beta_i u_k, i \in C_k) \approx \exp(-t_k/\tilde{\eta}_k),
\end{align*}
where we take $\beta_i=1$ if $i \in U_1$ and $\beta_i=\omega =\log(\phi_2)/\log(\phi_1)$ otherwise. Suppose there are $|C_k|$ variables in cluster $k$, $m$ of which are in $U_2$ and the balance in $U_2$. We exploited the exchangeability assumption by permuting variables: we assigned weight $\omega$ to $m_k$ of the $|C_k|$ in turn, for each of the $\binom{|C_k|}{m_k}$ combinations. We then computed the probability of exceedance as before, taking $\min_{i \in C_k} E_i/\beta_i$ as a structure variable and repeating the procedure for every permutation. We averaged the exceedance probabilities over all permutations; estimates are reported in \Cref{tbl:probsemipar}.

\begin{table}[htbt]
\centering
\caption{Probability estimates per cluster for the semiparametric approach based on the 0.985 threshold ($k=200$ largest observations)}
\label{tbl:probsemipar}
\begin{tabular}{lrrrrr}
\toprule
  & $C_1$ & $C_2$ & $C_3$ & $C_4$ & $C_5$\\
\midrule
$\log \widehat{p}_1$ & $ -6.76 $ & $ -16.55 $ & $ -9.22 $ & $ -6.81 $ & $ -17.01 $\\
$\log \widehat{p}_2$ & $ -7.06 $ & $ -18.82 $ & $ -10.83 $ & $ -6.93 $ & $ -17.82 $\\
\bottomrule
\end{tabular}
\end{table}

\subsection{Exchangeable models for asymptotically independent extremes}
\label{sec:heffernantawnexchangeable}
Since three clusters display strong evidence of asymptotic independence, we also considered the conditional extremes model described succinctly in \Cref{sec:heffernantawn}, this time under the assumption of strong pairwise extremal exchangeability \citep{Heffernan.Tawn:2004}. Data are mapped to unit Laplace margins, following \cite{Keef:2013b}, although all observations display positive dependence.

\subsubsection{Pseudo likelihood with skew-normal residuals}
Under strong exchangeability, the model of  \Cref{sec:heffernantawn} simplifies considerably since the conditional distributions of the residuals $Z_{k \mid j}$ and $Z_{j \mid k}$ $(j=1, \ldots, m; j \neq k)$ are equal for all pairs from the cluster of size $m$ and so are the parameters of the model \citep{Heffernan.Tawn:2004}. Preliminary exploration showed that the marginal distribution of the residuals from \cref{eq:resid} is leptokurtic and positively skewed. To account for this fact, we assume that residual components are conditionally independent and follow marginally a skew-normal distribution, with density $f_{\mathsf{SN}}(\cdot; \nu, \omega, \kappa)$, where $\nu$, $\omega$ and $\kappa$ are respectively location, scale and slant parameters.

Since the skew-normal distribution is a location-scale family, it follows
from the stochastic representation of the Heffernan--Tawn model
that
\begin{align*}
Y_k \mid Y_j > u \sim \mathsf{skewnormal}\left(\nu = \alpha Y_j + \mu Y_j^\beta, \omega =  \sigma Y_j^\beta, \kappa\right), \qquad k \in \{1, \ldots, m\} \setminus j.
\end{align*}
We estimate $(\alpha, \beta, \mu, \sigma, \kappa)$ by  maximizing the pseudo log likelihood
\begin{align*}
\sum_{j=1}^{m} \sum_{i: y_{ij} >u} \sum_{k \neq j}
\log f_{\textsf{SN}}(y_{ik}; \alpha y_{ij} + \mu y_{ij}^\beta, \sigma y_{ij}^\beta, \kappa).
\end{align*}
Only the parameters $\alpha \in [-1, 1]$ and $\beta \leq 1$, which characterize the asymptotic dependence, are of interest and we treat the other ones as nuisance parameters; parameter estimates are reported in \Cref{table:htestim} and we note that $\widehat{\alpha}=1$ for the clusters of asymptotically dependent variables.

The simulation approach outlined in \Cref{sec:heffernantawn}  yields estimates of the joint probability of exceedance which are exactly zero even with $10^{7}$ Monte Carlo replications for clusters of nearly independent variables with our data. The next section outlines an alternative strategy to palliate to this problem.

\subsubsection{Alternative estimation of the tail probability}
Consider a generic conditioning variable $Y_0$ exponentially distributed above $v$, meaning $Y_0 - v \mid Y_0 > v \sim \mathsf{Exp}(1)$. We assume that the conditional extremes model holds exactly and denote by $\boldsymbol{Z}$ the $m-1$ vector of residuals, its minimum entry by $Z^{\min}$ and the density of $Z^{\min}$ with support $\mathcal{Z}$ by $f_Z$. We write
\begin{align*}
 p &=\Pr(\alpha Y_0 + Y_0^\beta \boldsymbol{Z} > v \boldsymbol{1}_{m-1} \mid Y_0 > v) \Pr(Y_0 > v)\\&= \Pr\left(\left. Z^{\min} > \frac{v - \alpha Y_0}{ Y_0^\beta} \,\right|\, Y_0 > v\right)\Pr(Y_0 > v) \\
 &= \int_{\mathcal{Z}}\Pr\left(\left.\frac{v - \alpha Y_0}{Y_0^\beta}  < z   \,\right|\, Y_0 > v, Z^{\min}=z\right)\Pr(Y_0 > v) f_Z(z) \mathrm{d} z \\&= \int_{\mathcal{Z}}\Pr\left(Y_0   > v(z)  \mid Y_0 > v, Z^{\min}=z\right)\Pr(Y_0 > v) f_Z(z) \mathrm{d} z \\
 & = \int_{\mathcal{Z}} \exp[-\{v(z)-v\}]\Pr(Y_0 > v) f_Z(z) \mathrm{d} z
\end{align*}
Since $\alpha, \beta > 0$, both $y^\beta$ and $\alpha y$ are monotonically increasing functions of $y$ and the event of interest is equivalent to $Y_0 > v(z) > v$; the value of $v(z)$ can be found via root finding.

 Under strong pairwise extremal exchangeability, we can draw from the pool of residuals obtained by considering any exceedance, which gives on average $m$ times as many residuals to choose from as the original inferential approach. This decision is supported by tests of equality of distribution based on energy statistics \citep{Rizzo.Szekely:2010} for the residuals, obtained by taking the same parameters for all conditioning variables.

With $z^{\min}_i = \min_{j=1}^{m-1} z_{ij}$ for $i=1, \ldots, N$, we get the estimator
\begin{align*}
 \widehat{p} &= \frac{1}{N}\sum_{i=1}^{N} \exp\{-v(z^{\min}_i)\},\\
 \shortintertext{but, to avoid numerical overflow, we compute instead the log probability as}
 \log(\widehat{p}) &= -\min_{i=1}^N v(z^{\min}_i) - \log(N) + \log \left[ \sum_{i=1}^N \exp\left\{-v(z^{\min}_i) + \min_{i=1}^N v(z^{\min}_i)\right\}\right].
\end{align*}

We can proceed similarly for the second prediction task, where stations in group $U_1$ ($U_2)$ must exceed $s_1$ ($s_2$) and $s_1 > s_2$: we are after $p_2=\Pr(Y(s_i) > s_i, i=1, \ldots, 50)$. Since $s_2$ is lower than the threshold level, we could fit the conditional extremes model conditioning only on exceedances for station in area $U_1$, and then compute the minimum of the residual vector for each of groups $U_1$ and $U_2$ separately. This yields minimum values for $Y_0$, say $v_{1}(z^{\min, 1})$ and $v_{2}(z^{\min, 2})$, and it suffices to consider the probability that $Y_0$ exceeds the maximum of those two values. However, this approach does not ensure that the probability $p_2$ is less than $p_1$ and leads to a much lower number of residuals. Leveraging the exchangeability assumption again, we instead count how many variables appear in each of $U_1$ and $U_2$ for the cluster, consider permutations of the stations with the same number of variables in each group and repeat the procedure with all permutations of variables. We return the average probabilities in \Cref{table:htestim}.

\begin{table}[tbp!]
\centering
\caption{Conditional extremes modelling of Task~4 for different clusters  based on exceedances above the 0.98 quantile: cluster size, estimated probability of simultaneous exceedance (log scale) for events $S_1$ and $S_2$ and parameters estimates of the Heffernan-Tawn model.}

\begin{tabular}{lrrrrr}
\toprule
  & $C_1$ & $C_2$ & $C_3$ & $C_4$ & $C_5$\\
\midrule
cluster size & $8$ & $12$ & $13$ & $8$ & $9$\\
$\log(\widehat{p}_1)$ & $-6.153$ & $-18.243$ & $-8.502$ & $-6.352$ & $-26.723$\\
$\log(\widehat{p}_2)$ & $-6.517$ & $-20.888$ & $-10.370$ & $-6.514$ & $-27.675$\\
$(\widehat{\alpha}, \widehat{\beta})$ & $(1, 0.35)$ & $(0.099, 0.33)$ & $(0.3, 0.43)$ & $(1, 0.48)$ & $(0.098, 0.22)$\\
\bottomrule
\end{tabular}
\label{table:htestim}
\end{table}

\subsection{Joint tail exceedance for asymptotically dependent models}
\label{sec:mgp}
For clusters displaying evidence of asymptotic dependence, we fitted multivariate generalized Pareto distributions to threshold exceedances of logistic, negative logistic, H\"usler--Rei{\ss} and extremal Student type (both of the latter with exchangeable dependence structure). The probability of falling in the risk region can then be determined by considering the probability of any component exceeding the marginal threshold (which can be estimated empirically) and the probability that the multivariate generalized Pareto vector falls inside the risk region. While this could be obtained by simulating observations from the model, analytic expressions for the measure can be derived, as hinted in \cite{Kiriliouk:2019}.

Our starting point for this is
\cite{Dombry.Engelke.Oesting:2016}, who identified the distribution of the rescaled extremal function for the most popular parametric models employed in the literature. We compute the average intensity of the point process associated to the extreme value model over the set $\big\{\boldsymbol{Y} \in \mathbb{R}^D: \min_{j=1}^DY_j/u_j > 1 \big\}$ corresponding to joint exceedances for some generator vector $\boldsymbol{Y}$,
\begin{align*}
\varXi(\boldsymbol{u}) &= \int_{\mathbb{R}^D} \int_0^\infty \I{\zeta \min_{j=1}^D y_j/u_j > 1}
\zeta^{-2} \mathrm{d}\ \zeta f(\boldsymbol{y})\mathrm{d}\boldsymbol{y}
\\& = \int_{\mathbb{R}^D}\min_{j=1}^D \frac{y_j}{u_j} f(\boldsymbol{y})\mathrm{d}\boldsymbol{y}
\\&=  \sum_{j=1}^D\frac{1}{u_j} \int_{\mathbb{R}^D} y_j \I{y_ju_i/u_j < y_i, i =1, \ldots, D, i \neq
j} f(\boldsymbol{y})\mathrm{d}\boldsymbol{y}.
\end{align*}
We write $\varXi(\boldsymbol{u})=\sum_{j=1}^D \psi_j$ and use the terms of the weighted sum in \Cref{algo:composamp}.

These integrals are readily calculated for commonly employed parametric models.
For the logistic multivariate generalized Pareto model with parameter $\beta>1$,
\begin{align*}
 \varXi(\boldsymbol{u}) = \sum_{j=1}^D \frac{1}{u_j} \sum_{s \in \mathcal{P}(\{1, \ldots, D\} \setminus j}(-1)^{|s|}\left(1+\sum_{i \in s} k_{ij}^{-\beta}\right)^{1/\beta - 1}.
\end{align*}
The calculations are given in \Cref{{appendix:jointexceedanceprob}}, along with those for the negative logistic model with parameter $\theta>0$, for which
\begin{align*}
 \varXi(\boldsymbol{u}) =s_u^{-1/\theta-1} \sum_{j=1}^D u_j^{1/\theta}.
\end{align*}

For the Brown--Resnick or H\"usler--Rei{\ss} model, the proof mimicks \cite{Huser.Davison:2013} and we have for a $D \times D$ variogram matrix $\boldsymbol{\Gamma}$ that
\begin{align*}
   \varXi(\boldsymbol{u})&=\sum_{j=1}^D \frac{1}{u_j} \Phi_{D-1} \left\{\log(u_j\boldsymbol{1}_{D-1})-\log(\boldsymbol{u}_{-j}); \boldsymbol{\Gamma}_{-j,j},
\boldsymbol{\Sigma}_{-j}\right\},
\end{align*}
where $\boldsymbol{\Sigma}_{-j} = \boldsymbol{\Gamma}_{-j,j}\boldsymbol{1}_{D-1}^\top + \boldsymbol{1}_{D-1}\boldsymbol{\Gamma}_{j,-j}- \boldsymbol{\Gamma}_{-j,-j}$ and $\Phi_{k}(\cdot; \boldsymbol{\mu}, \boldsymbol{\Sigma})$ denotes the distribution function of a $k$ dimensional Gaussian vector with location $\boldsymbol{\mu}$ and scale $\boldsymbol{\Sigma}$.
The tail probabilities for the multivariate Gaussian can be efficiently estimated using the minimax exponential tilting (MET) estimator of \cite{Botev:2017}.
 We get likewise for the extremal Student-$t$ model with $\nu$ degrees of freedom and  correlation matrix $\boldsymbol{\Sigma}$ a weighted average of Student-$t$ distribution functions,
\begin{align*}
\varXi(\boldsymbol{u})&=\sum_{j=1}^D \frac{1}{u_i} \mathsf{St}_{D-1} \left\{ -\left(\frac{\boldsymbol{u}_{-j}}{u_j}\right)^{1/\nu}; -\boldsymbol{\Sigma}_{-j,j},
\frac{\boldsymbol{\Sigma}_{-j,-j} - \boldsymbol{\Sigma}_{-j,j}\boldsymbol{\Sigma}_{j,-j}}{\nu+1},\nu+1\right\},
\end{align*}
where $\mathsf{St}_{k}(\cdot, \boldsymbol{\mu}, \boldsymbol{\Sigma}, \nu)$ denotes the distribution function of a $k$ dimensional Student-$t$ distribution with location $\boldsymbol{\mu}$, scale $\boldsymbol{\Sigma}$ and $\nu$ degrees of freedom.

We estimated the parameters of the multivariate generalized Pareto distributions above a high threshold via maximum likelihood and computed the measure of the region of interest, which is $\varXi(\boldsymbol{s_1})/V(\boldsymbol{u})$. To obtain the joint probability of exceedance, we multiply the result by the empirical estimate of $\Pr(\max_{i=1}^D\{Y_i/u_i\} > 1)$, given by the proportion of points exceeding the threshold. The extremal Student and negative logistic were hard to fit and preliminary results showed poor performance relative to other parametric models, so we ignore them in the sequel.

For more complex models, we could resort to Monte Carlo methods to evaluate the probability of landing in the extreme region, say $\mathcal{R}$. For this, we need to be able to simulate points from the limiting Poisson point process measure over a region that comprises fully $\mathcal{R}$. The easiest option is to use the $R$-Pareto process associated with the sum risk functional, as this corresponds to a balanced mixture of extremal functions (i.e., the spectral density of the $\| \cdot\|_1$ norm) if we take equal thresholds \citep{Dombry.Engelke.Oesting:2016}.

To simulate observations from the limiting model over the risk region, we can do likewise and thin the point process. \cite{deFondeville:2018} uses such an accept-reject scheme for $R$-Pareto processes, but its efficiency decreases with the dimension of the problem. If the risk functional can be decomposed via indicators and linear combinations of variables (examples include weighted maxima, minima, averages and projections), one can directly simulate observations from a different $R$-Pareto process using the mixture representation. \cite{Ho.Dombry:2019} proposed such an approach for simulating from multivariate generalized Pareto Brown--Resnick vectors, but the procedure is more general and underexploited.

\Cref{algo:composamp} provides pseudo-code for a composition sampling algorithm. Note that neither normalizing constants $V(\boldsymbol{u})$ nor $\varXi(\boldsymbol{u})$ are needed to calculate the weights in \Cref{algo:composamp}, as we only need to know the weights up to proportionality for each variable. Second, we can easily bypass the analytical calculation of the weights, if the integrals were intractable, by simulating from the extremal functions and computing empirically the proportion of times a variable is the largest (or the smallest). For exchangeable models, these weights are uniform. Finally, the conditional simulations in the second step amount to univariate truncated distributions if the extremal functions are independent, but otherwise can be done efficiently for elliptical distributions.

\begin{algorithm}
 \caption{Composition sampling for standard $R$-Pareto vectors based on sum, min or max risk functionals}
\label{algo:composamp}
\begin{enumerate}
\item  Sample an index $I$ in $\{1, \ldots, D\}$ with probability
\begin{enumerate}[(a)]
\small
\itemsep0em
 \item max: $\Pr(I=j) = \varphi_j/V(\boldsymbol{u})$, where $\varphi_j = \frac{1}{u_j} \int_{\mathbb{R}^D} z_j \I{z_j \geq z_i, i \neq
j} f(\boldsymbol{z})\mathrm{d}\boldsymbol{z}$;
 \item min: $\Pr(I=j) = \psi_j/\varXi(\boldsymbol{u}_D)$, where $\psi_j = \frac{1}{u_j} \int_{\mathbb{R}^D} z_j \I{z_j \leq z_i, i \neq
j} f(\boldsymbol{z})\mathrm{d}\boldsymbol{z}$;
 \item sum: $\Pr(I=j) = 1/D$.
\end{enumerate}
 \item Sample extremal functions:
 \begin{enumerate}[(a)]
 \small
 \itemsep0em
 \item max: simulate a realization $Z_I$ from the $I$th marginal distribution of $P_I$, then draw truncated components from $\Pr(\boldsymbol{Z}_{-I} | \boldsymbol{Z}_{-I} \leq  Z_I)$;
 \item min: simulate a realization $Z_I$ from the $I$th marginal distribution of $P_I$, then draw truncated components from $\Pr(\boldsymbol{Z}_{-I} | \boldsymbol{Z}_{-I} \geq  Z_I)$;
  \item sum: simulate $\boldsymbol{Z} \sim P_{I}$.
 \end{enumerate}
\item Set $\boldsymbol{\omega} \gets \boldsymbol{Z}/Z_I$
\item  Simulate $R \sim \mathsf{Par}(1)$
\item Return  $Y \gets R\boldsymbol{\omega}$.
\end{enumerate}
\end{algorithm}

\begin{table}[tbp!]
\centering
\caption{Probability of simultaneous exceedance for clusters exhibiting asymptotic dependence based on multivariate generalized Pareto models fitted to exceedances above the 0.95 quantile and marginally censored below the median.}
 \begin{tabular}{lrrrr}
 \toprule
  & \multicolumn{2}{c}{H\"usler--Rei{\ss}} & \multicolumn{2}{c}{logistic} \\
  & $C_1$ & $C_4$ & $C_1$ & $C_4$   \\
  \midrule
coefficients & $0.567$ & $0.545$ & $0.594$ & $0.590$ \\
  $\log \widehat{p}_1$ & $-6.747$ & $-7.100$ & $-4.834$ & $-5.264$ \\
  $\log \widehat{p}_2$ & $-7.445$ & $-7.409$ & $-6.897$ & $-6.898$ \\
  $\widehat{\chi}$ & $0.451$ & $0.460$ & $0.490$ & $0.495$ \\
   \bottomrule
\end{tabular}
\end{table}

\subsection{Model selection} \label{sec:modselect}

It is difficult to assess the goodness-of-fit of extreme value models because there are few points in the region of interest. We could use information criteria to compare the different parametric models as in \cite{Kiriliouk:2019} for models fitted via maximum likelihood: in our example, the logistic model would be preferred over the H\"usler--Rei\ss{}, but no comparison with the Heffernan--Tawn model is possible.

Since we are interested in the joint probability of exceedance and data are assumed exchangeable, we consider an alternative cross-validation scheme for data from a cluster $C$ of $m$ variables, indexed $\{1, \ldots, m\}$. For each of the $n_k = \binom{m}{k}$ subsets of size $k$, denoted $S_i (i=1, \ldots, n_k)$, we compute the empirical estimator of $\widetilde{\chi}_{i,k}(u) = \Pr\{\min_{j \in S_i} F_j(X_j) > u\}/(1-u)$ at a high level. Using the parameter estimates obtained by fitting the model to the $m-k$ remaining variables in $C \setminus S_i$, we compute $\widehat{\chi}_{-i,k}$ based on the parametric model, $\chi = \Xi(\boldsymbol{1}_k; \boldsymbol{\theta}_{i})$, or via Monte Carlo simulations for the Heffernan--Tawn approach. We then compute the average $l_2$ distance,
\begin{align*}
l_2(u; k) = n_k^{-1}\left[\sum_{i=1}^{n_k} \left\{ \widetilde{\chi}_{i,k}(u) - \widehat{\chi}_{-i,k}(u)\right\}^2\right]^{1/2},
\end{align*}
as metric: smaller values indicate a better performance.

More interesting perhaps is comparing the performance in case of unequal probability level. Consider a pair of uniform random variables $F(Y_i) = U_1, F(Y_2) = U_2$ and exceedances $U_i > u_i (i=1, 2)$, where $u_1 < u_2$. We estimate the probability of joint exceedance given the maximum is above $t<u_2$ by
\begin{align}
\omega_2(u_1, u_2, t) = \frac{\Pr\{F(Y_1) > u_1, F(Y_2) > u_2\}}{\Pr\{\max_{i=1}^2 F(Y_i) > t\} }
\label{eq:unequalprob}
\end{align}
and compare empirical and model-based estimates. For the multivariate generalized Pareto distributions, $\omega_2(u_1, u_2, t)=\Xi(u_1, u_2)/V(t\boldsymbol{1}_2)$. For the conditional extremes model, we approximate the probability in \Cref{eq:unequalprob} by Monte Carlo, simulating  $Y_2 \sim \mathsf{Exp}(1) - \log(u_2)$ then $Y_1 = \alpha Y_2 + Y_2^\beta Z$, with $Z$ drawn from the empirical distribution of residuals.

\Cref{tab:tbl:gofasymptdep} gives the resulting estimates based on all permutations of pairs and triples. According to all metrics, the logistic model is preferred over both H\"usler--Rei\ss{} and Heffernan--Tawn model. All differences were statistically significant at the 1\% level.

\begin{table}[htbp!]
\caption{\label{tab:tbl:gofasymptdep}Weighted metrics ($\times 1000)$ for asymptotically dependent clusters for pairwise measure $l_2(0.99; 2)$ (pairs), triplewise $l_2(0.99; 3)$ (triples) and $w_2(0.96, 0.99, 0.97)$ (unequal).}
\centering
\begin{tabular}{lrrrrrr}
  \toprule
  & \multicolumn{3}{c}{cluster 1}& \multicolumn{3}{c}{cluster 4}\\\cmidrule(l){2-4} \cmidrule(l){5-7}Model &pairs  & triples  & unequal  & pairs  & triples  & unequal \\ \midrule
logistic & 6.4 & 4.2 & 6.4 & 9.2 & 7.3 & 9.2 \\
  Hüsler--Rei{\ss} & 12.3 & 16.0 & 19.3 & 16.6 & 19.1 & 17.8 \\
  Heffernan--Tawn & 11.5 & 8.3 & 61.3 & 14.6 & 10.7 & 61.7 \\
   \bottomrule
\end{tabular}

\end{table}

\subsection{Postmortem}
\label{sec:postmortemTask4}
We used the Heffernan--Tawn conditional extremes model for all clusters for the final submission. In hindsight, it turns out that choosing a much lower threshold in Task~4 leads to larger probabilities of joint exceedances and estimates closer to the truth, regardless of the estimation method employed.

\begin{figure}[htbp!]
\centering
\includegraphics[width=0.9\textwidth]{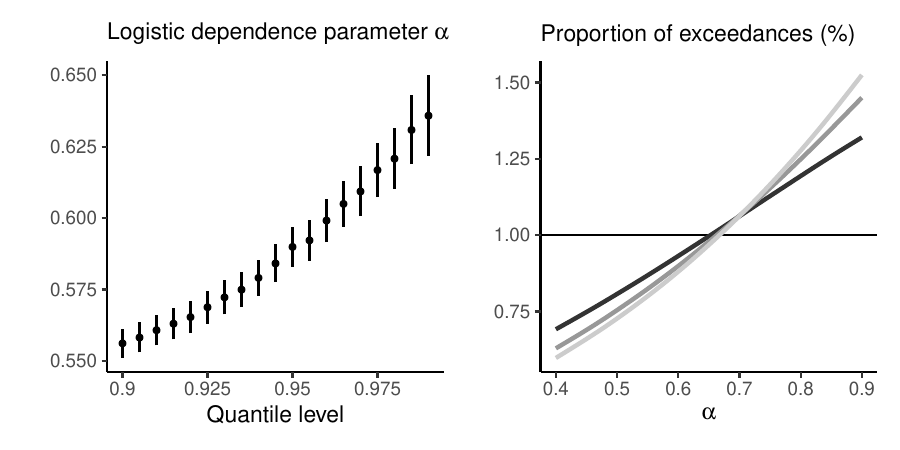}
 \caption{Left: threshold stability plot, with parameters estimates of the logistic with  95\% pointwise confidence interval for cluster $C_4$, estimated using censored likelihood. Right: proportion of sample exceedances exceeding the threshold $\max_{j=1}^8 Y_i > F^{-1}(q)$ from each of 100 mixtures components with parameter values $\alpha$ in $0.4$, for different threshold levels $q \in \{0.8, 0.9, 0.95\}$.}
\label{fig:threshstab}
\end{figure}

While the logistic model seemed better based on the metrics reported in \Cref{tab:tbl:gofasymptdep}, there was strong evidence of a lack of threshold stability for the logistic model, the property that underpins the extrapolation.  The left panel of \Cref{fig:threshstab} shows the estimated model parameter for the logistic multivariate generalized Pareto distribution as a function of the threshold, with observations censored below marginal medians. Under known margins, the mean square error of the maximum likelihood estimator $\widehat{\alpha}$ is proportional to $(nd)^{-1}$ \citep{Hofert:2012} as a result of exchangeability, so confidence intervals are unsurprisingly narrow. The plots suggest that dependence weakens at higher levels, while the conditional extremes model extrapolation (not shown) was much more stable.

A posteriori, it can be determined that this behaviour is a result of model misspecification, as the observations in the cluster were drawn from an infinite mixture of multivariate extreme value distributions, with $\alpha \sim \mathsf{U}(0.4, 0.9)$. The right panel of \Cref{fig:threshstab} shows results of an equibalanced mixture over a grid of 100 values between $0.4$ and $0.9$. We simulated $10^{7}$ samples from the logistic multivariate extreme value model and computed the proportion of time sample observations exceeded a certain quantile for each of the value of $\alpha$. We can see in \Cref{fig:threshstab} that higher values of $\alpha$ (corresponding to weaker dependence) occur in greater proportion, and more so as the threshold increases. This is in retrospective unsurprising since exceedances for multivariate generalized Pareto are defined in terms of marginal exceedance of $u$ in any component, $\{i: \max_{j=1}^8 Y_{ij} > u\}$. This observation has more general implications in applied data analysis, as many recorded environmental extremes can be viewed as the result of a mixture, e.g., a combination convective, cyclonic and orographic rainfall for precipitation extremes.

We note in passing that the teams with the best performance on Task~4 all used very low threshold, set to values around the 0.8 quantile. Based on the previous result, it seems that higher threshold levels would have lead to more acute preferential sampling problems for methods based on methods that condition on at least one variable being large.

\section{Discussion}
The 2023 edition of the Extreme Value Analysis data challenge was unusual in that it featured multiple tasks, with simulated data rather than real data. Given the nature of the challenges, most of the approaches employed by the different teams partaking in the competition were firmly rooted in the extreme value analysis literature, with rather fewer machine learning approaches than in previous editions.   All tasks required accurate point estimates, and adequate uncertainty quantification was only essential for Task~1  to obtain good coverage. Multiple teams used off-the-shelf methods implemented in \textbf{R} packages, or the methodology developed in their own institution to tackle the problems. Software availability and complexity of bespoke implementation are still a barrier and more so for practitioners not necessarily well versed into the intricacies of the different models.

What did we learn from partaking in the competition? For one thing, it is (and remains) hard to validate models for extremes and the uncertainty of estimates is often so large that it isn't clear how useful the results are. In Task~1, we used a scoring rule to discriminate between models and this suggested that simpler generalized Pareto models were better. All models had indistinguishable coverage due to the high variability, and the postmortem analysis showed that neither of the metrics was particularly trustworthy at the higher return levels sought.

Sometimes, cutting corners short could still lead to a good performance. It was possible to ignore the covariates completely in Task~2 and get an answer that would have ranked very close to the true value, highlighting the fact that simpler naive alternatives could perform decently even if they are not justified. We noticed after the competition was completed that different imputations of missing data yielded point estimates in Task~2 that were sometimes 10 to 20 units higher, highlighting the sensitivity of estimators at very high levels to lower-level components despite the fact that the two sets of imputed values were strongly correlated.  Using only complete cases to build the threshold and exceedance models would have been a valid strategy, but reduces the available sample size. Extrapolation is quite sensitive to small changes to parameters and models.

As a small team, time constraints forced us to make pragmatic rather than principled choices in order to obtain answers by the deadline. For example, we did not have time to reconsider Task~3 and did not incorporate covariates for the models, leading to a drop in ranking from 1st for initial submission to 6th in the final ranking. Exploratory analysis for the Coputopia data did not reveal dependence of extremes on covariates, and we note that there is a shortage of diagnostics to detect such dependence and test for the latter.
For models fitted via maximum likelihood, we can perform model selection using information criteria or hypothesis tests between nested models, but power to detect such changes will be necessarily limited and there is no such tool for semiparametric models. There is likewise a shortage of methods for goodness-of-fit assessment, and this is of course due to the nature of rare event modelling.

Multivariate models for extremes seldom extend to high dimensions, and their usefulness when they do is limited by their parametrization, with models that are overly simple or complex. There are still few useful and tractable  (semi)parametric models that can be fitted in high dimensions. The  conditional extremes model has a number of parameters that grows linearly with the dimension $d$ of the random vector if we condition on a single variable, whereas the H\"usler--Rei\ss{} model for asymptotically dependent extremes has $\mathrm{O}(d^2)$ parameters and can characterize pairwise dependence. Simplifying assumptions like exchangeability are unlikely to hold in any practical circumstance. Software implementation is also lagging behind.

We used multivariate generalized Pareto models for the asymptotically dependent clusters of Task~4, but we could have considered more general $R$-Pareto vectors with different risk functionals \citep{deFondeville:2018}. While in principle the limiting measure of the point process is the same regardless of the choice of risk functional $R$, in practice different functionals (whether it be the maximum of at least $k$ components, the minimum, the sum) leads to different samples and often very different parameter estimates. This ties in with the preferential sampling problem discussed in \Cref{sec:postmortemTask4}. Oftentimes, the return levels that are required through regulation are so high that the extrapolation is dubious, no matter the model, and the uncertainty is so large that the relevance of asking for such risk estimates is questionable. Naive Monte Carlo methods showed their limitations in Task~4, and this begs the question of whether development of new simulation algorithms for rare events, using exponential tilting or otherwise, would be necessary to obtain estimates in very rare instances. In Task~4, competing state-of-the-art methods returned estimates that were sometimes an order of magnitude different and this can be critical when designing infrastructure or policy regulations.

Our team tried to aggressively leverage information provided by the authors of the challenge \citep{rohrbeck2023editorial}, including the known marginal distributions for Tasks~3 and 4 and the fact that data were missing completely at random (MCAR) for the Utopia data. Most of the time, environmental data are missing because of extreme events, and preferential sampling of station location makes the MCAR assumption unlikely. Other assumptions that we made based on visual exploration helped simplify models to a large extent and get more precise estimates, although not all were true: independence turned out to be an incorrect working assumption, but exchangeability within clusters indeed held in Task~4. Choosing lower threshold models across the board would have given much more information than those constraints, but it is always difficult to select thresholds apriori.

Outside of the comparisons between different approaches for imputation of missing data and threshold modelling schemes, our contributions include the cross-validation scheme of \Cref{algo:intervalscore} and the weighted $l_2$ diagnostic of \Cref{sec:modselect}, which exploits the exchangeability by resampling variables rather than observations. The metrics we proposed in \Cref{sec:modselect} exploit structure of the models and could be adapted to the non-exchangeable setting by resampling observations rather than variables. The challenge then is that there are few exceedances and threshold stability is needed for validation. We also proposed an alternative tail estimation scheme in \Cref{sec:heffernantawnexchangeable}, along with the use of skew-normal distribution for the residuals of the Heffernan--Tawn. The formulae for the joint probability of exceedance for four parametric multivariate models derived in \Cref{sec:mgp}, in addition to the composition sampling algorithm in \Cref{algo:composamp}, can be used more broadly for modelling asymptotically dependent data using multivariate generalized Pareto distributions.

 Real-life applications come with a plethora of other challenges (including, but not limited to, nonstationarity, trends, mixtures, changes in distributions, etc.) that would have further complexified the tasks, so the situations and approaches considered in this paper remain somewhat utopic.

\section*{Statements and Declarations}
\subsection*{Acknowledgment} LRB wishes to thank Samuel Perreault for providing code to run exchangeability tests and for valuable insights regarding their implementation. The authors of the paper express gratitude to Christian Rohrbeck, Emma Simpson, and Jonathan Tawn for organizing the data competition.
\subsubsection*{Funding}
LRB acknowledges financial support from IVADO and the
Canada First Research Excellence Fund (FRG-2019-7771647733), and from the Natural Sciences and Engineering Research Council of Canada (RGPIN-2022-05001). AH acknowledges financial support from the Indian Institute of Technology Kanpur and Rice University collaborative research grant under Award No. DOIR/2023246.
\subsubsection*{Availability of data and materials}
Data are available from the \href{https://bocconi-my.sharepoint.com/:f:/g/personal/stefano_rizzelli2_unibocconi_it/ErEDfDLGMUdGmGn2$V_2$K24jQBbZXxJIzU6aZ7UR6X_davbw?e=eMfeFY}{conference website}.
\subsubsection*{Conflict of interest}
The authors declare that they have no conflict of interest.
\subsubsection*{Code availability}
Reproducible \textbf{R} code to generate all figures and tables is provided in an online repository at \href{https://github.com/lbelzile/EVA2023-data-challenge}{\texttt{https://github.com/lbelzile/EVA2023-data-challenge}}.
\subsubsection*{Authors' contributions}
All authors participated in discussions and writing. RY worked on Tasks 1--3, LB on Tasks 2 and 4 and AH lead the work on Task~1.

\bibliographystyle{apalike2}
 {\small
\bibliography{EVA_challenge.bib}
}

\appendix
\section{Simultaneous exceedance for multivariate generalized Pareto vectors}
\label{appendix:jointexceedanceprob}

\subsection{Logistic model}
For the logistic distribution, we consider independent and identically distributed Fréchet generators with shape $
\beta>1$ and scale $c_\beta = \Gamma(1-1/\beta)^{-1}$ \citep{Dombry.Engelke.Oesting:2016}. If $k_{ij} = u_i/u_j$ and we denote by $\mathcal{P}(\cdot)$ the power set of a given vector of indices, then
\begin{align*}
 \varXi(\boldsymbol{u}) &=\sum_{j=1}^D \int_{\mathbb{R}^D_{+}} \I{y_i > k_{ij}y_j, i \neq j}\frac{y_j}{u_j} \prod_{k=1}^D \beta c_\beta^\beta y_k^{-1-\beta} \exp \left\{ - (y_k/c_\beta)^{-\beta}\right\} \d y_1 \cdots \d y_D
 \\&= \sum_{j=1}^D \frac{1}{u_j} \int_{0}^\infty \beta c_\beta^\beta y_j^{-\beta} \exp \left\{ - (y_j/c_\beta)^{-\beta}\right\}\prod_{\substack{i=1\\i \neq j}}^D\left[ 1- \exp\left\{- (y_jk_{ij}/c_\beta)^{-\beta}\right\}\right] \d y_j
 \\&= \sum_{j=1}^D \frac{1}{u_j} \int_{0}^\infty \beta c_\beta^\beta y_j^{-\beta} \sum_{s \in \mathcal{P}(\{1, \ldots, D\} \setminus j)}(-1)^{|s|}   \exp\left\{- y_j^{-\beta} \left(1+\sum_{i \in s} k_{ij}^{-\beta}\right)c_\beta^{\beta}\right\} \d y_j
 \\&=  \sum_{j=1}^D \frac{1}{u_j} \sum_{s \in \mathcal{P}(\{1, \ldots, D\} \setminus j}(-1)^{|s|} \int_{0}^\infty c_\beta^\beta x_j^{-1/\beta} \exp\left\{- x_j \left(1+\sum_{i \in s} k_{ij}^{-\beta}\right) c_\beta^{\beta}\right\} \d x_j \\
 &= \sum_{j=1}^D \frac{1}{u_j} \sum_{s \in \mathcal{P}(\{1, \ldots, D\} \setminus j}(-1)^{|s|}\left(1+\sum_{i \in s} k_{ij}^{-\beta}\right)^{1/\beta - 1}
\end{align*}
where the penultimate step follows from the change of variable $x_j = y_j^{-\beta}$ and from integrating the unnormalized density of a gamma distribution. If $\boldsymbol{u} = u\boldsymbol{1}_D$, the measure simplifies to
\begin{align*}
 \varXi(u \boldsymbol{1}_D) = \frac{D}{u} \sum_{k=0}^{D-1} \binom{D-1}{k} (-1)^k (1+k)^{1/\beta - 1}
\end{align*}

\subsection{Negative logistic model}
We can proceed likewise with the negative logistic model with parameter $\theta>0$, whose generator is Weibull with shape $\theta$ and scale $c_{\theta}=1/\Gamma(1+1/\theta)$. Write $s_u = \sum_{k=1}^D u_k$; we have
\begin{align*}
 \varXi(\boldsymbol{u}) & = \sum_{j=1}^D \int_{\mathbb{R}^D_{+}} \I{y_i > k_{ij} y_j, i \neq j}\frac{y_j}{u_j} \prod_{k=1}^D \theta c_\theta^{-\theta}y_j^{\theta-1} \exp \left\{-\left(y_i/c_{\theta}\right)^\theta\right\}\d y_1 \cdots \d y_D
  \\ & = \sum_{j=1}^D \frac{1}{u_j} \int_0^\infty \theta c_{\theta}^{-\theta}y_j^{\theta} \exp \left\{-(s_u/u_j)\left( y_j/c_{\theta}\right)^\theta\right\}\d y_j\\
  &=s_u^{-1/\theta-1} \sum_{j=1}^D u_j^{1/\theta}
\end{align*}
where the last integral is the expectation of a Weibull distribution with scale $c_{\theta}(s_u/u_j)^{-1/\theta}$ and shape $\theta$.

\end{document}